\documentstyle[epsf,psfig]{article}
\usepackage{cite}

\def\z{\zeta}                                   
\def\k{\kappa}                                  %\def\Psi{\Psi}

\def\z{\zeta}                                   

\def\k{\kappa}                                   %\def\Psi{\Psi}

\def\p{\pi}                      
                    
\def\s{\sigma}                
\def\t{\tau}

\def\f{\frac}
\def\pr{\partial}
\def\eq{equation}

\begin{document}

%\jl{11}

\begin{center}
%\title
{\LARGE \bf On the properties of two pulses propagating\\
\noindent
simultaneously in different dispersion\\
\vspace{3mm}
\noindent
regimes in a nonlinear planar waveguide}\\
\end{center}
\vspace{5mm}
%\author
{\large{\bf Monika E. Pietrzyk}
\vspace{5mm}
\footnote{On leave from: Faculty of Physics, 
Warsaw University of Technology, Warsaw, Poland
%Institute of Fundamental 
%Technological Research, Polish Academy of Sciences, Warsaw, Poland.
%Visiting Scientist at the Abdus Salam International Centre for 
%Theoretical Physics, 34-100 Trieste, April - October 1998.
}}\\
%\address
{Institut f\"ur Festk\"orpertheorie und Theoretische Optik,
Friedrich-Schiller \\
\noindent
Universit\"at Jena, D-07743, Jena, Germany.}

\vspace{5mm}

\noindent
%\begin{abstract}
\begin{minipage}{12cm}
Properties of two pulses propagating simultaneously in different dispersion 
regimes, anomalous and normal, in a Kerr-type planar waveguide are 
studied in the framework of the nonlinear Schr\"odinger equation. 
It is found that the presence of the pulse propagating in a normal 
dispersion regime can cause termination of catastrophic self-focusing 
of the pulse with anomalous dispersion. It is also shown that the coupling 
between pulses can give rise to spatio-temporal splitting of the pulse 
propagating in anomalous dispersion regime, but it does not necessarily 
lead to catastrophic self-focusing of the pulse with normal dispersion.
For the limiting case when the dispersive term of the pulse 
propagating in normal dispersion regime can be neglected an indication 
(based on the variational estimation) of a possibility of a stable 
self-trapped propagation of both pulses is obtained. This stabilization 
is similar to the one which was found earlier in media 
with saturation-type nonlinearity. \vspace{2mm} \\ 
%\end{abstract}
 \vspace{2mm} 
\end{minipage}
%\maketitle

%\pagebreak

\section {Introduction}
The propagation of a dispersive light pulse in a planar  
waveguide with positive, instantaneous Kerr-type nonlinearity can be 
described by the (2+1)-dimensional nonlinear Schr\"odinger equation
(NSE) \cite{Silberberg:coo}: 
\begin{equation}
i \f {\pr }{\pr \z} \Psi + \f {\s} 2 \f{\pr^2}{\pr \t^2} \Psi + 
\f 12 \f{\pr^2}{\pr \xi^2}\Psi + |\Psi|^2 \Psi = 0,   
\label{nse}
\end{equation}
where the parameters $\z, \t, \xi$ are as defined in appendix A. 
 
Equation (\ref{nse}) is valid only for pulses in the picosecond range; 
for shorter pulses additional terms, due to a higher-order dispersion, 
for example,  should be included. The last term in equation (\ref{nse}) 
describes Kerr-type nonlinearity; second and third terms are associated, 
respectively, with diffraction, which causes spreading of the pulse in 
space, and first-order group velocity dispersion, which leads to temporal 
broadening of the pulse. Parameter $\sigma$, which can be either positive 
(for anomalous dispersion) or negative (for normal dispersion), is the 
dispersion-to-diffraction ratio \cite{Ryan:pca}. The spatio-temporal 
dynamics of the pulse depends, to a high degree, on the sign of this 
parameter. 
 
It is known that some solutions of the (2+1)-dimensional NSE can develop 
into a singularity of the electric field in the self-focus point. 
This phenomenon, known as catastrophic self-focusing, occurs 
simultaneously in space and time for pulses propagating in planar 
waveguides with a\-no\-ma\-lo\-us group velocity dispersion 
(equation (\ref{nse}) with $\s > 0$) 
\cite{Silberberg:coo, Desaix:vatc}, and also for dispersionless beams 
propagating in self-focusing bulk media (equation (\ref{nse}) with the 
dispersive term replaced by a diffraction term) \cite{Zakharov:eto}
when parameters of the system are above the threshold of catastrophic
self-focusing \cite{Fibich:sbn}, which is usulally computed with 
the aid of the method of moments \cite{Vlasov:, Rasmussen:bin, Cao:sfo},
the variational method \cite{Desaix:vatc, Manassah:sfs}, 
and also numerical simulations \cite{Silberberg:coo, Chernev:, Pietrzyk:nso}.
The occuurence of catastrophic self-focusing is not only non-physical, 
it also prevents examination of the pulse behavior behind the self-focus, 
for it emerges just as an artifact  of
approximations made when deriving the NSE. In order to avoid this  
limitation, either some nonlinear stabilization mechanisms 
such as saturation \cite{Karlsson:obi} or non-locality \cite{Suter:} 
of nonlinearity, Raman scattering \cite{Dyshko:mso}, plasma formation 
\cite{Yablonovitch:aia}, multiphoton ionization \cite{Henz:tds}, 
higher-order group velocity dispersion terms \cite{Karpman:sos}, 
an adequate composition of the above mentioned effects \cite{Blair:(2+1)}, 
or a non-paraxial treatment of the process of self-focusing 
\cite{Akhmediev:dtn, Fibich:sbn, Fibich:sfi, Gurwich:ana, 
Feit:bnf,Sheppard:nst} should be included into consideration. 
However the standard paraxial NSE can still serve as the model 
equation for self-focusing in the case when parameters of the system
are below the threshold of catastrophic self-focusing or, in the reverse case,
for studying dynamics of a pulse/beam in the prefocal region. 

Another situation occurs when the pulse propagates in a normal dispersion 
regime. In this case the terms describing dispersion and diffraction have 
different signs and two different effects, spatial self-focusing and 
temporal self-defocusing, simultaneously influencing the propagation 
of the pulse. This causes the situation where, in the solution of the 
NSE (equation (1) with $\s < 0$) neither singularity \cite{Berkshire:cit} 
nor localized steady-states occurs \cite{Litvak:anw}. Moreover, this solution 
is accompanied by a breaking of spatio-temporal symmetry and a uniform 
structure of the pulse and can finally lead to an occurrence of several 
humps in the field distribution \cite{Litvak:anw}, splitting of the 
pulses into two sub-pulses \cite{Pietrzyk:pik}, or splitting into several 
sub-pulses \cite{Gross:eos}. It has also been reported that the presence 
of even very small normal dispersion can lead to the destruction of 
soliton breathers propagating in nonlinear planar waveguides
\cite{Burak:sbs}. In the case of the (3+1)-dimensional NSE splitting of a 
pulse into two sub-pulses has also been observed 
\cite{Rothenberg:stf, Chernev:sfo, Luther:sft, Rothenberg:psd}, while
splitting into several sub-pulses predicted theoretically in 
\cite{Fibich:sfi,Trippenbach:dos,Zozulya:pdo,Zozulya:dos,Ranka:bot}
has been confirmed experimantally by the authors of
\cite{Fibich:sfi, Gross:eos, Zozulya:pdo, 
Ranka:bot, Zozulya:dos, Ranka:oop, Diddams:aap}.

Thus, depending on the sign of dispersion, a dispersive pulse propagating 
in a Kerr-type planar waveguide reveals different behaviour. Catastrophic 
self-focusing (in the framework of the NSE) takes place in the case 
of anomalous dispersion. For normal 
dispersion the typical process is spatio-temporal splitting. It seems 
interesting to study an interaction between two pulses co-propagating
in such a medium, i.e. a Kerr-type planar waveguide, 
under the assumption that one of them propagates in a normal 
dispersion regime and another is in an anomalous regime. To the 
author's knowledge this problem has not been studied in the literature and 
the main purpose of this paper is to consider it. Note that 
interaction of spatially separated light beams whose evolution is modeled
by a set of $n$ ($n \ge 2$) nonlinearly coupled NSEs 
was studied by several authors \cite{McKinstrie:nfo, Berge:cai, 
Desyatnikov:iot, Bang:fca}. Moreover, the importance of the interaction 
between two pulses in a nonlinear medium has been pointed out already 
by Agrawal in \cite{Agrawal:ifo}, where an intriguing effect of an 
induced focusing of two beams co-propagating in a self-defocusing 
medium has been reported.

It is also known that neither for anomalous dispersion \cite{Konno:sfo}
nor for normal dispersion \cite{Litvak:anw} do stable soliton-like 
solutions of the (2+1)-dimensional  (and also (3+1)-dimensional) 
NSE exist. This statement also concerns experimental results, since 
no soliton-like solution has been observed in pure Kerr-like 
nonlinear media with two or three transverse dimensions. From the point 
of view of applications, i.g. as elements of optical switching devices 
\cite{McLeod:(3+1)}, the existence of stable soliton-like solutions is very
important. Therefore, solutions to this problem has been already proposed by
several authors: for example, it has been shown that soliton-like
structures can be realized in media with saturation-type nonlinearity 
\cite{Karlsson:obi, Vidal:ebb, gatz:poo, Skarka:ssp}, in photorefractive 
media \cite{Segev:ssi, Shih:tds}, in media with quadratic nonlinearity
\cite{Malomed:ssi, Berge:fos, Hayata:msi, Mihalache:sws, Agin:got, 
Torruellas:oot}, in media with cascaded $\chi^{(2)} - \chi^{(3)}$ 
nonlinearity \cite{Liu:goo, Berge:sfa}, and also in the limiting case of 
the discrete-continuous NSE which can model propagation of short optical 
pulses in an array of linearly coupled optical fibers \cite{Darmanyan:ioc}. 
In this paper we will consider another possibility of obtaining a 
self-trapped solution in two transverse dimensions, namely in a 
configuration of the (1+1)-dimensional NSE coupled to 
the (2+1)-dimensional NSE. 

We proceed as follows. In section 2, two coupled 
NSEs describing the co-propagation of two dispersive pulses in a nonlinear 
planar waveguide and basic equations following from the variational 
method will be introduced. Next, in section 3, the problem of 
catastrophic self-focusing will be considered. First, the influence of the 
parameters of the pulse propagating in a normal dispersion regime on the 
threshold of catastrophic self-focusing of the pulse propagating in 
an anomalous dispersion regime will be studied. We will also examine 
whether catastrophic self-focusing of the pulse propagating in a normal 
dispersion regime can occur as a result of the nonlinear coupling 
between two pulses. In section 4, which is devoted to the problem of 
spatio-temporal splitting, we will investigate whether the influence of 
the pulse propagating in a normal dispersion 
regime can enforce spatio-temporal splitting of the pulse with anomalous 
dispersion. In the last section, section 5, we will focus on the 
limiting case when the dispersive term of the normal pulse can be neglected. 
In this case the problem of two coupled (2+1)-dimensional NSEs will be 
reduced to the system of a (1+1)-dimensional NSE coupled to a 
(2+1)-dimensional NSE. 
The main reason to study this configuration is to investigate a possibility 
of a stable, self-trapped solution. 

The interaction between pulses will be assumed to be limited to 
cross-phase modulation, a nonlinear effect through which the phase of an 
optical beam/pulse is affected by another propagating beam/pulse and 
which can cause a redistribution of energy within each be\-am/pul\-se. 
Another effect, four-wave mixing,  will be neglected, so that no energy 
transfer between both pulses will be taken into consideration.  
The analysis presented in this paper is based on the variational method
\cite{Anderson:vatn} and numerical simulations using the split-step spectral 
method \cite{Agrawal:nfo}. 
Throughout the paper the pulse propagating in an anomalous (normal)
dispersion regime will be referred to as the {\it anomalous} 
({\it normal}) pulse.  
 
\section {Basic equations}
The co-propagation of two optical pulses in a nonlinear planar waveguide 
can be described by two coupled nonlinear Schr\"odinger equations:
$$ i \frac {\partial} {\partial \zeta} \Psi_1 + \frac {\sigma_1} 2 
\frac{\pr ^2}
{\pr \tau  ^2} \Psi_1 + \frac 1 2 \frac {\pr ^2} 
{ \partial \xi ^2} \Psi_1 + 
(\mid \Psi_1 \mid^2 + 2 \mid \Psi_2 \mid^2 )\Psi_1 =0 , 
\refstepcounter{equation} \eqno (\theequation a) \label{nsetwo}$$
$$i \frac {\partial} {\partial \zeta} \Psi_2 
+ \frac {\sigma_2} 2 \frac{\partial ^2}
{\partial \tau  ^2} \Psi_2 + \mu \frac 1 2 \frac {\partial ^2} 
{ \partial \xi ^2} \Psi_2 + r (\mid \Psi_2 \mid^2 
+ 2 \mid \Psi_1 \mid^2 )\Psi_2 =0,
\label{2nse} 
\eqno (\theequation b)$$
where the last terms represent cross-phase modulation,
a nonlinear effect which causes a coupling between pulses 
and the terms before the last ones describe self-phase modulation.

It is assumed that the subscript $j=1$ $(j=2)$ denotes the anomalous (normal) 
pulse, hence  $\s_1 > 0$ and $\s_2 < 0$. The notations in equations 
(\ref{2nse}a,b) are explained  in appendix A. 
The initial conditions will be taken in the form of the Gaussian pulses 
\begin{\eq}
{\Psi_j}(\z=0, \t, \xi) =\sqrt{\k_j}
exp{\left[- \f 1 2 \t^2 \left( 1+iC_{\t j}\right) \right]}
exp{\left[- \f 1 2 \xi^2 \left( 1+iC_{\xi j}\right) \right]},
\label{gauss}
\end{\eq}
where $C_{\t j}$ $(C_{\xi j})$ is  
the temporal (spatial) chirp of the $j$-th pulse, $j=1,2$. 
The parameter $\k_j$ will be called here the strength of nonlinearity 
of the j-th pulse (see explanation in appendix A).
 
\subsection {Variational method}
It is known that the set of NSEs (equation (\ref{2nse}a,b)) 
can be obtained from the Lagrangian density given by
$$
L=\f {i} 2 \left(\Psi_1^* \f {\pr \Psi_1} {\pr \z} 
- \Psi_1 \f {\pr \Psi_1^*} {\pr \z}\right) 
+ \f {i} 2 \f 1 r \left(\Psi_2^* \f {\pr \Psi_2} {\pr \z} 
- \Psi_2 \f {\pr \Psi_2^*} {\pr \z}\right)
$$
\begin{\eq}
-\f 1 2 \left|\f {\pr \Psi_1}{\pr \xi} \right| ^2
-\f {\s_1} 2 \left| \f {\pr \Psi_1}{\pr \t} \right| ^2 
-\f 1 2 \f {\mu} r \left| \frac {\pr \Psi_2}{\pr \xi} \right| ^2
-\f 1 2 \f {\sigma_1} r \left| \frac {\pr \Psi_2}{\pr \t} \right| ^2
\label{lagrtwo}
\end{\eq}
$$
+\f 1 2 \mid \Psi_1 \mid ^4 +2 \mid \Psi_1 \mid ^2 \mid \Psi_2 \mid ^2
+\f 1 2 \mid \Psi_2 \mid ^4.
$$
Following the variational method \cite{Anderson:vatn} let us choose 
a proper multi-parametric trial function for the solution of 
equation (\ref{2nse}a,b). Since in this paper we consider the Gaussian 
initial condition (equation (\ref{gauss})) it is natural to take as 
the trial function the Gaussian function: 
\begin{\eq}
%\fl 
\Psi_j=A_j(\zeta)exp{\left[- \f 1 2 \f{\tau^2}{ w_{\tau j(\zeta)}}\right]} 
exp{\left[- \f 1 2 \f{\xi^2} {w_{\xi j}(\zeta)} \right]}
exp{\left[ \f {i} 2 {\tau^2 C_{\t j}(\z)} \right]} 
exp{\left[\f {i} 2 {\xi^2 C_{\xi j}(\z)} \right]},
\label{ansatz}
\end{\eq}
with 12 parameters: the complex conjugate amplitudes, 
$A_j, A^*_j$, the temporal and the spatial widths, $w_{\t j}, w_{\xi j}$,
and the temporal and the spatial chirps, $C_{\t j}, C_{\xi j}$, where $j=1,2$. 
{}From the initial condition (equation (\ref{gauss})) it follows that
$A_j(\z=0)=\sqrt{\k_j}$, $w_{\t j}(\z=0)=w_{\xi j}(\z=0)=1$.

The evolution equations for the parameters of the trial function
are obtained by varying the reduced Lagrangian
$$
\langle L \rangle:=\int \limits_{-\infty}^{\infty} L d\xi d\tau, 
$$
into which the trial function (equation (\ref{ansatz})) is inserted, 
with respect to the parameters of the trial function, 
$A_j, A^*_j$, $w_{\t j}$, $w_{\xi j}$, $C_{\t j}$, $C_{\xi j}$. 
We obtain the following 12 coupled ordinary differential equations: 
$$ \frac {d} {d z} {\cal I}_1=0
\refstepcounter{equation} \eqno (\theequation a) \label{inten}
$$
$$\frac {d} {d z} {\cal I}_2=0
\eqno (\theequation b)
$$
$$\frac {d^2 w_{\tau 1}}{d \zeta ^2} =
\frac {\sigma _1^2}{w_{\tau 1}^3}
- \frac {\sigma_1} 2 \frac {{\cal I}_1}{w_{\tau 1}^2 w_{\xi 1}}
- \frac {4 {\cal I}_2 w_{\tau 1} \sigma_1} 
{(w_{\tau 1}^2 +w_{\tau 2}^2)^{\frac 3 2} 
(w_{\xi 1}^2 + w_{\xi 2}^2)^{\frac 1 2 }}
\refstepcounter{equation} \eqno (\theequation a) $$
$$\frac {d^2 w_{\xi 1}} {d \zeta ^2} =
\frac 1 {w_{\xi 1}^3}
- \frac 1 2 \frac {{\cal I}_1}{w_{\tau 1} w_{\xi 1}^2}
- \frac {4 {\cal I}_2 w_{\xi 1}} 
{(w_{\tau 1}^2 +w_{\tau 2}^2)^{\frac 1 2} 
(w_{\xi 1}^2 + w_{\xi 2}^2)^{\frac 3 2 }}\eqno (\theequation b) $$
$$\frac {d^2 w_{\tau 2}} {d \zeta ^2} =
\frac {\sigma_2^2}{w_{\tau 2}^3}
- \frac {\sigma_2} 2 \frac {{\cal I}_2 r}{w_{\tau 2}^2 w_{\xi 2}}
- \frac {4 {\cal I}_1 w_{\tau 2} \sigma_2 r} 
{(w_{\tau 1}^2 +w_{\tau 2}^2)^{\frac 3 2} 
(w_{\xi 1}^2 + w_{\xi 2}^2)^{\frac 1 2 }}\eqno (\theequation c) $$
$$\frac {d^2 w_{\xi 2}} {d \zeta ^2} =
\frac {\mu ^2} {w_{\xi 2}^3}
- \frac {\mu} 2 \frac {{\cal I}_2 r}{w_{\tau 2} w_{\xi 2}^2}
- \frac {4 {\cal I}_1 w_{\xi 2} \mu r} 
{(w_{\tau 1}^2 +w_{\tau 2}^2)^{\frac 1 2} 
(w_{\xi 1}^2 + w_{\xi 2}^2)^{\frac 3 2 }}
\label{width}
\eqno (\theequation d)$$
$${C_{\tau 1}}=-\frac 1 {\sigma_1} \frac {d \, ln(w_{\tau 1})}{d z}
\refstepcounter{equation} \eqno (\theequation a)$$
$${C_{\xi 1}}=-\frac {d \, ln(w_{\xi 1})}{d z}\eqno (\theequation b)$$
$${C_{\tau 2}}=-\frac 1 {\sigma_2} \frac {d \, ln(w_{\tau 2})}{d z}
\eqno (\theequation c)$$
$${C_{\xi 2}}=-\frac 1 {\mu} \frac {d \, ln(w_{\xi 2})}{d z}
\eqno (\theequation d)$$
$$\frac {d \phi_1}{d z}=\frac 3 4 |A_1|^2 
- \frac 1 2 \frac {\sigma_1}{w_{\tau 1}^2} + \frac 1 {w_{\xi 1}^2}
+ {\cal I}_1 \frac {2 + {w_{\tau 1}^2}/{(w_{\tau 1}^2+w_{\tau 2}^2)}
+ {w_{\xi 1}^2}/{(w_{\xi 1}^2+w_{\xi 2}^2)}}
{(w_{\tau 1}^2+w_{\tau 2}^2)^{\f 1 2} (w_{\xi 1}^2+w_{\xi 2}^2)^{\f 1 2}} 
\refstepcounter{equation} \eqno (\theequation a)$$
$$\frac {d \phi_2}{d z}=\frac 3 4 r | A_2 |^2 
- \frac 1 2 \frac {\sigma_2}{w_{\tau 1}^2}+ \frac {\mu} {w_{\xi 2}^2}
+{\cal I}_2 r \frac {2 + {w_{\tau 2}^2}/{(w_{\tau 1}^2+w_{\tau 2}^2)}
+{w_{\xi 2}^2}/{(w_{\xi 1}^2+w_{\xi 2}^2})}
{(w_{\tau 1}^2+w_{\tau 2}^2)^{\f 1 2} (w_{\xi 1}^2+w_{\xi 2}^2)^{\f 1 2}} 
\label{phase}
\eqno (\theequation b)$$
where
${\cal I}_j:= w_{\t j}(\z)w_{\xi j}(\z)|A(\z)|^2=\k_j=const$. From equations 
(6a) and (6b), which are actually the energy conservation 
laws for two pulses, $N_j:=\int \int |\Psi_j|^2 d\t d\xi = \p {\cal I}_j$,  
it follows that there is no energy transfer between the pulses.

The set of equations (6a)-(9b) is rather complicated and only 
in the special case when $\s_1=\s_2=1, {\cal I}_2=0$ 
is the analytical solution 
\[
w_{\xi j}(\z)=w_{\t j}(\z)=
\left[1+\z^2 \left(1-\f{\k_j}2\right)\right]^{\f 1 2}, \hspace{3mm} j=1,2
\] 
available \cite{Anderson:}. More general situations should be 
treated numerically, e.g. using  the Runge-Kutta method \cite{fortran}. 
Still, equations (7a)-(7d) can be simplified to one evolution
equation. To proceed, let us first rewrite equations (7a)-(7d)
in the form
$$
\f {d^2 w_{\t 1}}{d \z ^2} = -\f {\s_1} 2 \f {\pr}{\pr w_{\t 1}}
V_1 (w_{\t 1} ,w_{\xi 1}, w_{\t 2}, w_{\xi 2}),\hspace*{4mm}\\
\f {d^2 w_{\xi 1}}{d \z ^2} = -\f 1 2 \f {\pr}{\pr w_{\xi 1}}
V_1 (w_{\t 1} ,w_{\xi 1}, w_{\t 2}, w_{\xi 2}),$$
$$\f {d^2 w_{\t 2}}{d \z ^2} = -\f {\s_2} 2 \f {\pr}{\pr w_{\t 2}}
V_1 (w_{\t 1} ,w_{\xi 1}, w_{\t 2}, w_{\xi 2}),\hspace*{4mm}\\
\f {d^2 w_{\xi 2}}{d \z ^2} = -\f {\mu} 2 \f {\pr}{\pr w_{\xi 2}}
V_1 (w_{\t 1} ,w_{\xi 1}, w_{\t 2}, w_{\xi 2}),
$$
where the potentials $V_1(w_{\t 1} ,w_{\xi 1}, w_{\t 2}, w_{\xi 2})$
and $V_2(w_{\t 1} ,w_{\xi 1}, w_{\t 2}, w_{\xi 2})$ read as
$$V_1(w_{\t 1} ,w_{\xi 1}, w_{\t 2}, w_{\xi 2}):=
\f {\s_1}{w^2_{\t 1}} + \f 1 {w^2_{\xi 1}} 
- \f {{\cal I}_1}{w_{\t 1} w_{\xi 1}} - \f {4 {\cal I}_2} 
{(w^2_{\t 1} + w^2_{\t 2})^{\f 1 2} (w^2_{\xi 1} + w^2_{\xi 2})^{\f 1 2}},$$
$$V_2(w_{\t 1} ,w_{\xi 1}, w_{\t 2}, w_{\xi 2}):=
\f {\s_2}{w^2_{\t 2}} + \f {\mu} {w^2_{\xi 2}} 
- \f {{\cal I}_2 r}{w_{\t 2} w_{\xi 2}} - \f {4 {\cal I}_1 r} 
{(w^2_{\t 1} + w^2_{\t 2})^{\f 1 2} (w^2_{\xi 1} + w^2_{\xi 2})^{\f 1 2}}.$$

It can be also shown that the quantity
$W := r {\cal I}_1 W_1 +{\cal I}_2 W_2 $, where
$$W_1 := \f 1 {\s_1} \left( \f {d w_{\t 1}}{d \z} \right)^2 +  
\left( \f {d w_{\xi 1}}{d \z} \right)^2 + V_1, \hspace*{5mm}\\
W_2 := \f 1 {\s_2} \left( \f {d w_{\t 2}}{d \z} \right)^2 +  
\f 1 {\mu} \left( \f {d w_{\xi 2}}{d \z} \right)^2 + V_2$$ 
is a constant of motion.

Again, using equations (7a)-(7d) it can be calculated that 
\begin{equation}
\f {d^2 \bar w} {d \z^2} = 2 W,
\label{aver}
\end{equation}
where $\bar w:= r {\cal I}_1 \left(\f {w^2_{\t 1}}{\s_1} 
+ w^2_{\xi 1}\right) + {\cal I}_2 \left(\f {w^2_{\t 2}}{\s_2}
+ \f {w^2_{\xi 2}}{\mu} \right)$ (here we assume 
$\s_1 \neq 0,\s_2 \neq 0, \mu \neq 0$). 

From equation (\ref{aver}) one can easily get the evolution equation 
for $\bar w$
\begin{\eq}
\bar w (\z) = W \z^2 + \z \left. \f{\pr \bar w}{ \pr \z} \right|_{\z=0}
+ \bar w(\z=0).
\label{evol}
\end{\eq} 

\section{Catastrophic self-focusing}
This section is devoted to the problem of catastrophic self-focusing, 
which can occur in the solution of the set of equation (\ref{2nse}a,b). 
Our analysis is based on the variational method and
numerical simulations and the comparison of the results of both.  
Note that once we have specified the threshold of catastrophic self-focusing 
we then know for which parameters of the system the NSE is valid and we 
can use this information in further research. 

From the point of view of analytical estimations, which can be done 
using the method of moments \cite{Cao:sfo, Cornolti:egb} or the variational 
method \cite{Desaix:vatt} catastrophic self-focusing is identified with 
a development  of a singularity in the solution at a finite distance of 
propagation. 

Let us briefly discuss the case of a single pulse, i.e., let us
make the assumption that ${\cal I}_2=0$ and
${\cal I} := {\cal I}_1, \s :=\s_1, w_{\t} := w_{\t 1}, w_{\xi} := w_{\xi 1}$.
Then we get that
$$W := \f 1 {\s} \left(\f {d w_{\t}}{d \z} \right)^2   
+ \left(\f {d w_{\xi}}{d \z} \right)^2 +V,$$ with the potential 
$$V(w_{\t}, w_{\xi}) := \f {\s}{w^2_{\t}} +  \f 1 {w^2_{\xi}} - 
\f {I} {w_{\t} w_{\xi}}$$ and 
$$\bar w = \f {w^2_{\t}} {\s}+ w^2_{\xi},$$
whereas equations (\ref{aver}) and (\ref{evol}) still remain valid. 
From equation (\ref{evol}) it follows that the quantity $\bar w$ goes
to zero on a finite distance of propagation when one of the following
conditions is satisfied
\addtocounter{equation}{1}
$$\left\{ \begin{array}{ll}
\vspace*{2mm} 
W<0,\\ 
\vspace*{2mm}
W=0 \hspace{2mm} \hbox{and} \hspace{2mm} 
\left. \f {\pr \bar w}{\pr \z} \right|_{\z=0} < 0,\\
\vspace*{2mm}
W>0 \hspace{2mm} \hbox{and} \hspace{2mm} 
\left. \f {\pr \bar w}{\pr \z} \right|_{\z=0} \leq -2 \sqrt{W  \bar w(0)}.
\label{cond}
\end{array} \right. \eqno({\theequation }) $$

A vanishing of $\bar w$ can be associated with a singularity of the solution
of the NSE (equation (\ref{nse})) only when dispersion is anomalous, 
$\s>0$, since only in this case the quantity $\bar w $ can be interpreted 
as an average width of the pulse and the condition $\bar w =0$ is equivalent
to a simultaneous vanishing of both widths of the pulse. Therefore, for
the Gaussian initial condition (equation (\ref{gauss}) with $\k := \k_1$)
without initial chirp, $C_{\t}(0) := C_{\t 1}(0) = 0, 
C_{\xi}(0) := C_{\xi 1}(0) = 0$, i.e. for 
$\left. \f {\pr \bar w}{\pr \z} \right|_{\z=0} =0$, catastrophic self-focusing
of the pulse with anomalous dispersion will arise when the condition
$W < 0$ is satisfied, i.e. when 
\begin{equation} 
\k > \k_{V cat} = \s + 1.
\label{onecoll}
\end{equation}
Note that the condition given by equation (\ref{onecoll}) 
agrees with the results obtained
in \cite{Cao:sfo} with the aid of the method of moments 
for an elliptic Gaussian beam.

Another situation occurs in the case of normal dispersion, namely 
a vanishing of the quantity $\bar w$ means only that 
$w^2_{\t 1} =\s w^2_{\xi}$, therefore, nothing about catastrophic 
self-focusing can be concluded from equation (\ref{evol}). 
However, based on the method of moments it has been demonstrated that 
catastrophic self-focusing in this case does not occur 
\cite{Berkshire:cit}. 

In our numerical simulations catastrophic self-focusing is identified 
with a discontinuity of the phase $\phi(\t,\xi,\z)$ of the amplitude 
$\Psi:=|\Psi| e^{i \phi}$ in the central point of the coordinate system, 
$\t=0, \xi=0$, and with non-monotonic behavior of the intensity $|\Psi|^2$ 
in the central point after catastrophic self-focusing has been reached 
\cite{Pietrzyk:nso}. The threshold of catastrophic self-focusing given 
by the numerical analysis \cite{Chernev:, Chernev:sfo, Pietrzyk:nso}
\[ \k_{N cat} \approx \s + 0.885\] is lower than the one given by analytical 
estimations.  

Let us examine now two coupled NSEs given by equations (\ref{2nse}a), 
(\ref{2nse}b). We can consider three different cases: (i) both pulses 
propagate in an anomalous dispersion regime, $\s_1 > 0, \s_2 > 0$; 
(ii) both pulses propagate in normal dispersion regime, $\s_1 < 0, \s_2< 0$; 
(iii) pulses propagate in different dispersion regimes, anomalous and normal,
$\s_1 > 0, \s_2< 0$.

In the first case, when both pulses propagate in anomalous dispersion 
regimes the threshold of catastrophic self-focusing can be calculated
in a similar way as it was done for a single pulse and  
is given by equation (12). For the Gaussian initial condition 
(equation (\ref{gauss})) without initial chirp, 
$C_{\t 1}(0) = C_{\xi 1}(0) = C_{\t 2}(0) = C_{\xi 2}(0) = 0$, 
i.e. when $\left. \f {\pr \bar w}{\pr \z} \right|_{\z=0} = 0$,
catastrophic self-focusing occurs when the condition $W<0$, which
reads as 
\begin{\eq}
r {\cal I}_1 \left(\s_1 + 1 - \k_1 - 2 \k_2\right) 
+{\cal I}_2 \left( \s_2 + \mu - r \k_2 - 2 r \k_1 \right) < 0,
\label{twocoll} 
\end{\eq}
is satisfied.
 
Since vanishing of the quantity $\bar w$, which can be 
interpreted as an average width of the pulses, is associated with
a simultaneous vanishing of both widths of both pulses, then it can be
concluded that when catastrophic self-focusing of one of the 
pulse occurs, it also occurs for the second one. This conclusion and 
the condition (\ref{twocoll}) written for the symmetric case, 
$\s_1=\s_2=\mu=r =1$, agree with the results obtained in 
\cite{McKinstrie:nfo, Berge:cai, Desyatnikov:iot, Bang:fca} 
for two cylindrically symmetric, spatially separated beams whose 
distance vanishes. 

In the case of two pulses propagating in a normal dispersion regime 
the situation is simple: catastrophic self-focusing does
not develop, even for very large strengths of nonlinearity of the pulses.

The situation in more complicated when the pulses propagate in 
different dispersion regimes, anomalous and normal: the threshold
of catastrophic self-focusing cannot be calculated from equation 
(\ref{evol}), therefore numerical solutions of equations (7a)-(7d)
should be performed in order to analyse this problem. The first goal of our 
study is to examine an influence of the parameters of the normal pulse
on the threshold of catastrophic self-focusing of the anomalous pulse.  
The parameters of the anomalous pulse have, therefore, been chosen 
in such a way that the relations $\k_1 > \k_{V cat} = 1+\s_1$ 
(in the variational method), and $\k_1 > \k_{N cat} \approx 0.885+\s_1$ 
(in the numerical simulations) are satisfied, 
which mean that catastrophic self-focusing of the anomalous pulse
will develop when there is no coupling between pulses. Then the 
parameters of the normal pulse, i.e. the strength of nonlinearity, 
$\k_2$, and the dispersion-to-diffraction ratio, $\s_2$, are varied. 
We found that catastrophic self-focusing of the pulse propagating 
in an anomalous dispersion regime can be arrested by the 
influence of the pulse propagating in a normal dispersion regime. 

\begin{figure}
\raisebox{-2.5cm}[1.0cm][1cm]{
a)}\\
\hspace*{1cm}\raisebox{-2cm}[4cm][2cm]{
\psfig{figure=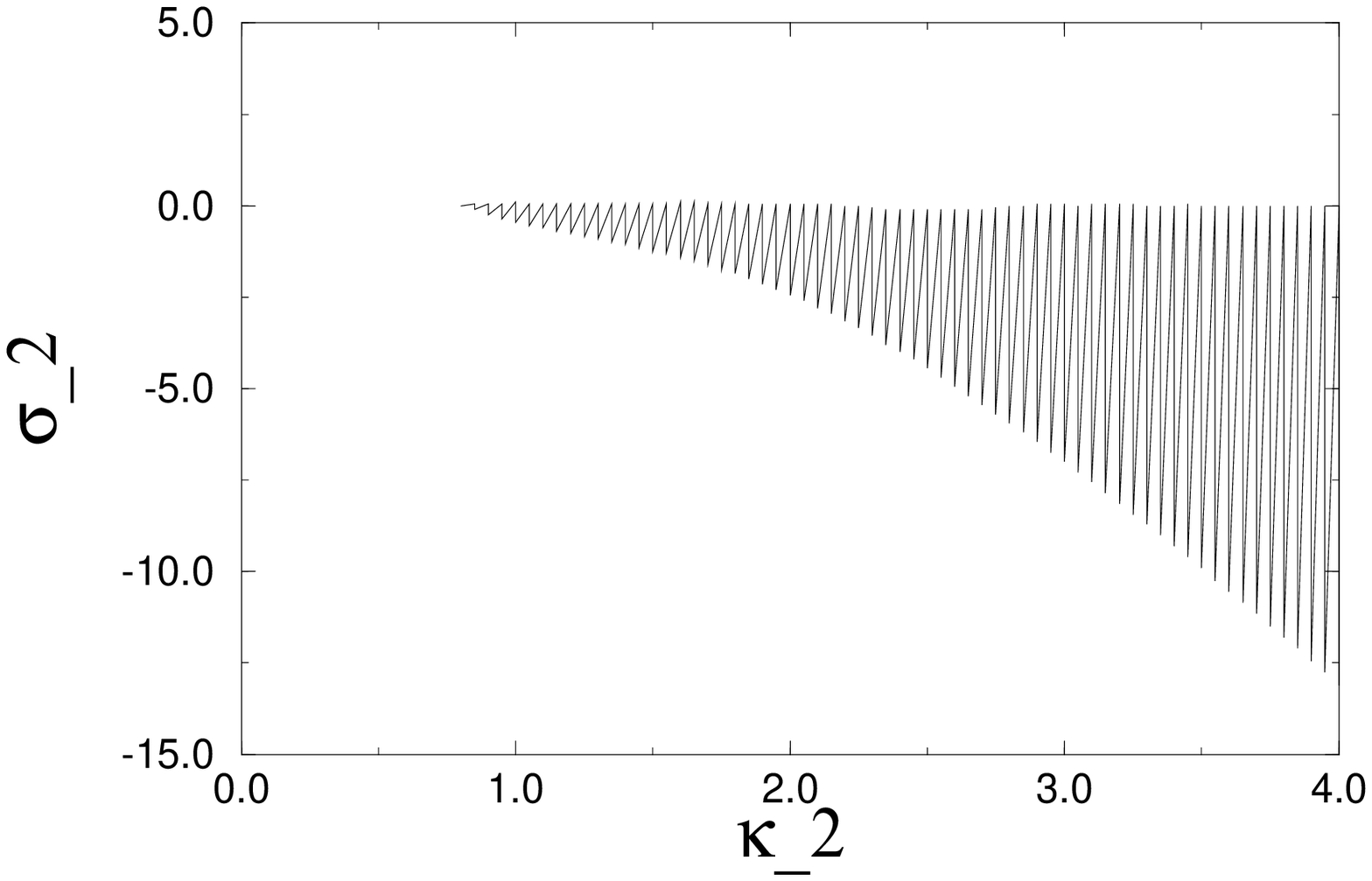,width=13cm}}\\
\raisebox{-2.5cm}[1cm][1cm]{
b)}\\
\hspace*{1cm}\raisebox{-2cm}[4cm][2cm]{
\psfig{figure=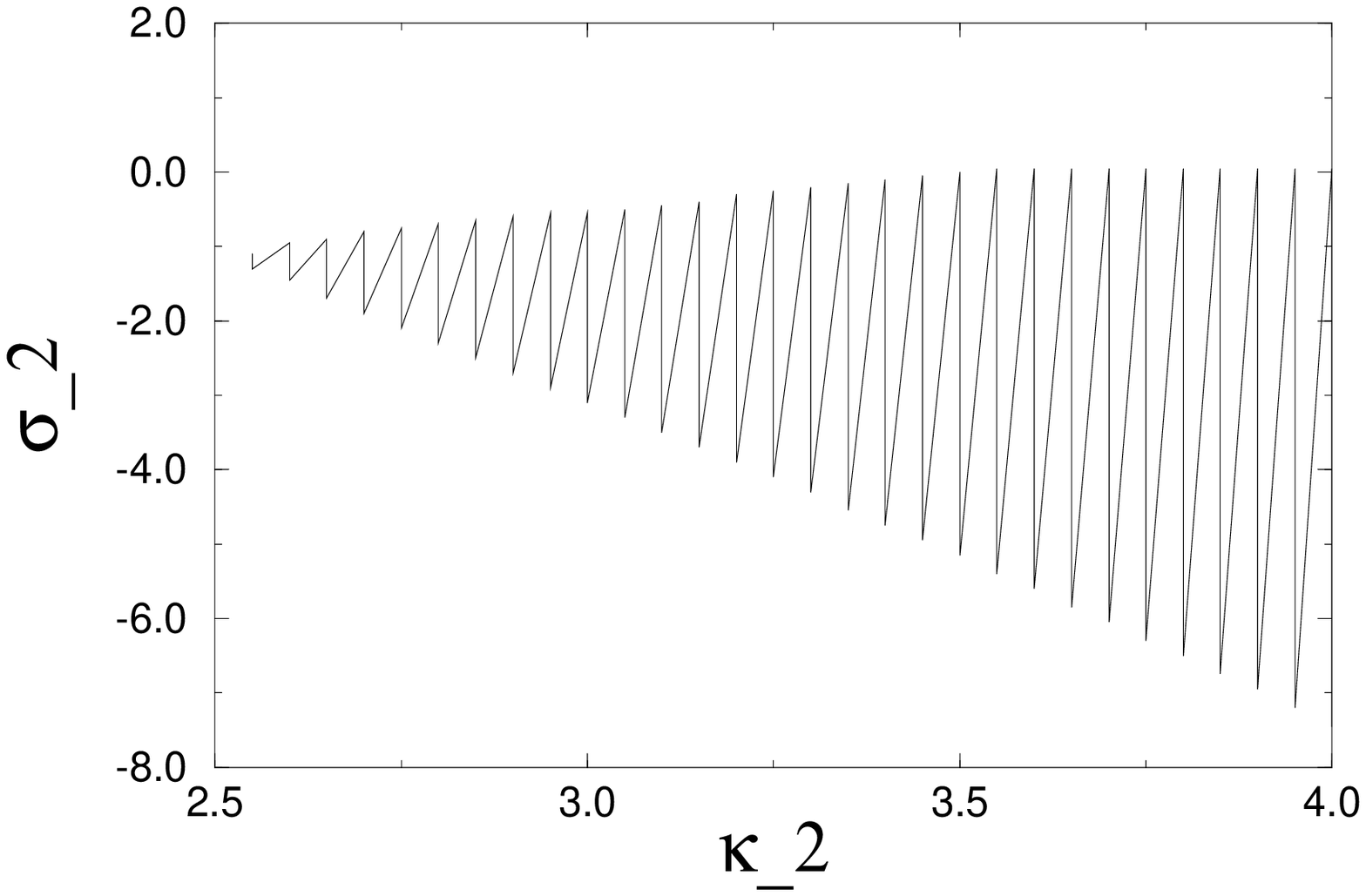,width=13cm}}
\caption{The results of the variational method
displaying the dependence of the threshold of catastrophic 
self-focusing of the pulse propagating in an anomalous dispersion regime,
$\Psi_1$,  on the parameters of the pulse propagating in a normal dispersion 
regime, $\Psi_2$. The shaded area denotes the range of the parameters,
the strength of nonlinearity, $\k_2$, and the dispersion-to-diffraction 
ratio, $\s_2$, for which catastrophic self-focusing does not occur. 
The parameters of the anomalous pulse have been chosen in such 
a way that they are above the threshold of catastrophic self-focusing
in a single propagation regime: i.e. for (a) $\k_1=2.2, \s_1=1.0$, 
for (b) $\k_1=2.3, \s_1=1.0$.}
\label{figvar}
\end{figure}

The results following from the variational method are shown in figure 
\ref{figvar}. The shaded area denotes the range of the parameters of 
the normal pulse, $\k_2$ and $\s_2$, for which catastrophic self-focusing 
of the anomalous pulse does not occur. It is evident that for small 
nonlinearity of the normal pulse, $\k_2$, the term describing cross-phase 
modulation of the anomalous pulse is negligible as  compared  with 
self-phase modulation. Therefore, the process of catastrophic 
self-focusing cannot be stopped and it takes place for all values of 
$\s_2$. When the strength of nonlinearity $\k_2$ increases, the influence 
of the normal pulse on the anomalous pulse through the cross-phase 
modulation term increases and then it is possible, for some values of 
the dispersion-to-diffraction ratio, 
$|\s_{V low} (\k_2)| <|\s_2| < |\s_{V upp} (\k_2)|$, 
to stop catastrophic self-focusing. 
The lower threshold, $|\s_{V low}(\k_2)|$, in the beginning decreases with an 
increase of the strength of nonlinearity of the normal pulse, $\k_2$. 
For sufficiently large nonlinearity, $\k_2 > \k_{V low}$, the lower 
threshold becomes zero. The upper threshold, $|\s_{V upp}(\k_2)|$, increases 
with an increase of nonlinearity. The existence of the lower threshold 
can be explained as follows: when $|\s_2| < |\s_{V low}|$ is small, the 
dispersive term of the normal pulse is negligible as compared with 
diffraction. Therefore, the most important role in the propagation 
of the normal pulse is played by self-focusing ,which 
not only does not lead to an arresting of catastrophic self-focusing, 
but even additionally enhances it. A similar 
situation is known, for example, in a configuration of two beams, 
which co-propagate in a bulk medium and have the same amplitudes 
\cite{McKinstrie:nfo}: namely the critical value of nonlinearity necessary 
for catastrophic self-focusing is three times smaller than in the case 
when they propagate as single pulses, as can be calculated i.g. from
equation (\ref{twocoll}). On the other hand, while for 
large $|\s_2|$, we have a broadening of the normal pulse with a significant 
spreading of energy out from the centre of the coordinate system, 
$\xi=0, \t=0$, while for the anomalous pulse there is a tendency of energy 
to concentrate in the centre. Then the overlap of two pulses becomes 
negligible, so that the coupling between them through cross-phase 
modulation is very small and catastrophic self-focusing of the anomalous 
pulse can not be stopped by the influence of the normal pulse. 

\begin{figure}
\hspace*{1cm}\raisebox{-3.5cm}[4.5cm][3cm]{
\psfig{figure=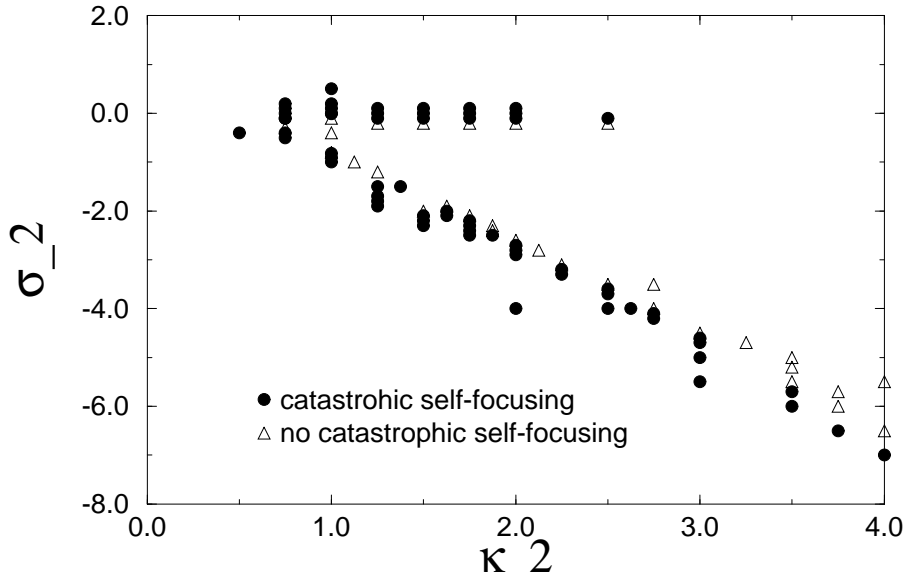,width=13cm}}
\caption{The results of the numerical simulations displaying 
the dependence of the threshold of catastrophic 
self-focusing of the pulse propagating in an anomalous dispersion regime,
$\Psi_1$,  on the parameters of the pulse propagating in a normal dispersion 
regime, $\Psi_2$, i.e. the strength of nonlinearity, $\k_2$, and
the dispersion-to-diffraction ratio $\s_2$. Full circle (empty triangle) 
points denote the occurrence (lack) of catastrophic self-focusing.     
The parameters of the anomalous pulse have been chosen in such 
a way that they are above the threshold of catastrophic self-focusing
in a single propagation regime, i.e. $\k_1=2.0, \s_1=1.0$.}
\label{fignum}
\end{figure}

The results obtained with the aid of the numerical calculations are shown 
in figure \ref{fignum}. They confirm predictions of the variational method. 
Namely, catastrophic self-focusing of the anomalous pulse can be arrested by 
the pulses propagating in a normal dispersion regime when the strength 
of nonlinearity is sufficiently large, $\k_2 > \k_{N low}$, and
the dispersion-to-diffraction ratio satisfies the relation 
$|\s_{N low}(\k_2)| < |\s_2| < |\s_{N upp}(\k_2)|$. 

Another question is whether the nonlinear coupling between pulses 
can cause catastrophic self-focusing of the pulse propagating in a normal 
dispersion regime. As it has been already mentioned, 
in the case of two simultaneously propagating pulses with anomalous 
dispersion catastrophic self-focusing of one of the pulse
is associated with catastrophic self-focusing of the other one, and
both widths of both pulses go to zero simultaneously when catastrophic
self-focusing occurs. 

However, the variational method demonstrates that in the case discussed 
here catastrophic self-focusing of the anomalous pulse does not necessarily 
lead to catastrophic self-focusing of the normal pulse. Namely, when 
catastrophic self-focusing of the anomalous pulse occurs, the normal 
pulse can demonstrate, depending on the parameters of the system, 
two different characteristics: (i) both widths of the pulse initially 
decrease reaching a minimum on a certain distance of propagation 
and then they start to increase; (ii) the spatial width of the pulse 
vanishes to zero on a finite distance of propagation whereas the
temporal width initially decreases, reaching a minimum on a certain 
distance of propagation, and then it increases. In particular, the case (i)
can be realized for the following parameters of the system 
$\k_1=3, \k_2=3, \s_1=1, \s_2=-7$, while the case (ii) occurs, 
for example, when $\k_1=3, \k_2=3, \s_1=1, \s_2=-1$. 
An effect, similar to (ii) has been also 
observed in the case of a pulse that propagates in a bulk medium with
normal dispersion and whose dynamics is modeled by the 
(3+1)-dimensional NSE: namely catastrophic self-focusing of this pulse
occurs when only the spatial widths vanish to zero while the 
temporal width is not allowed to reach zero value \cite{Berge:sfo}.

The numerical simulations have not confirmed the results
of the variational method concerning a possibility of catastrophic 
self-focusing of the normal pulse whose spatial width vanishes to zero
on a cartain distance of propagation, $\z$, and whose 
temporal width is left larger than zero. However, no definite
statement that this effect is prohibited can be made either. 
Additional calculations should be performed to clarify this question. 

We can therefore conclude, based on the variational method and the numerical 
simulations, that catastrophic self-focusing of the anomalous pulse 
does not necessarily lead to catastrophic self-focusing of the normal pulse.

\section{Spatio-temporal splitting}
In this section the problem of spatio-temporal splitting is discussed 
in more detail. The origin of spatio-temporal splitting of a single pulse 
propagating in a normal dispersion regime in Kerr-type planar waveguides 
\cite{Gross:eos, Burak:veo, Pietrzyk:pik} or bulk media 
\cite{Rothenberg:stf, Chernev:sfo, Luther:sft, Rothenberg:psd}
is the fact that spatial self-focusing (in one or two dimensions) 
and temporal self-defocusing act simultaneously during the propagation. 
Therefore, in space, there is a tendency of energy to concentrate 
in the centre of the coordinate system, $\t=0, \xi=0$, whereas in time 
a spreading of energy away from the centre takes place. When both 
effects are combined, local focusing areas develop away from the centre 
and, as a result, spatio-temporal splitting of the pulse into several 
sub-pulses takes place \cite{Gross:eos}. The number of sub-pulses 
emerging in this way, for a sufficiently large propagation distance, 
is proportional, as it has been proposed in \cite{Gross:eos}, 
to the order of the temporal soliton, $N := \sqrt {{\k}/{\s}}$, where
the parameters of the (2+1)-dimensional NSE (equation (\ref{nse}))
$\s$ and $\k$ denote respectivelly the dispersion-to-diffraction ratio
and the strength of nonlinearity. Specifically, splitting of the pulse
into two sub-pulses has been observed \cite{Pietrzyk:pik} in 
the system with parameters $\z= 2, \k= 4, \s= -0.1$, while in 
\cite{Gross:eos} splitting of the pulse into three sub-pulses 
has been obtained for $\z=0.15 , \k=100, \s= -3$. 

Although we have not verified the statement that the number of sub-pulses
is proportional to the order of a temporal soliton, $N$, since it was not 
the purpose of our study, we have observed that in the case of a single 
pulse with normal dispersion the tendency of the pulse 
to split increases when the strength of nonlinearity, $\k$, increases
and when the dispersion-to-diffraction ratio, $\s$, decreases. 
Still, it remains for us an open question 
whether the humps in the filed distribution of a pulse mentioned
in \cite{Litvak:anw} can be identified with sub-pulses whose existence 
is demonstrated in this paper and in \cite{Gross:eos, Pietrzyk:pik}. 
If this is not the case, we could take the opportunity to speculate that
the splitting of a pulse could not be observed by the authors of 
\cite{Litvak:anw} since the strength of nonlinearity used by them was 
relatively weak, $\k \approx 1.76$ (while the dispersion-to-diffraction 
ration was chosen to be $\s \approx -0.32$) and only small local humps, 
instead of full pulse splitting, could be detected. 

Some agreement between numerical and variational solutions of the NSE 
with normal dis\-per\-sion 
has been demonstrated in the literature. For example, 
in \cite{Litvak:anw, Gross:eos} it has been shown, respectively, 
that in both methods the number of oscillations of the peak amplitude 
of the pulse is the same and there is a similarity  in evolution of the 
average square widths of the pulses. Nevertheless, in all cases when 
spatio-temporal splitting of pulses has been observed, numerical 
simulations have been used \cite{Gross:eos, Chernev:sfo, Rothenberg:psd, 
Ranka:oop, Burak:veo}. The variational method is not appropriate to 
predict the splitting of pulses, since it requires the solution to have a 
shape which does  not change in propagation. When a Gaussian function is 
chosen as the initial condition, as we have done in this paper, 
it is difficult (if not impossible) to guess a trial function 
function which would satisfy the initial condition and also 
could describe spatio-temporal splitting of the pulse. Here, it is  
worthwhile to recall that the variational method cannot 
be applied to predict, for example, the formation of higher-order 
solitons in planar waveguides or optical fibers \cite{Anderson:vatn}. 

\begin{figure}
\raisebox{1.cm}[2.5cm][0cm]{
\hspace{4.0cm}$|\Psi_1|^2$}\\ 
\raisebox{-1.5cm}[0cm][0cm]{
$\hspace{4.8cm}\t$}\\
\raisebox{0.5cm}[0cm][0cm]{
\hspace{1.5cm}a)}\\
\raisebox{-1.5cm}[0cm][2cm]{
\hspace{3.5cm}\hspace{5.1cm}$\xi$}\\
\raisebox{2cm}[2.0cm][1cm]{
\hspace*{3.5cm}\psfig{figure=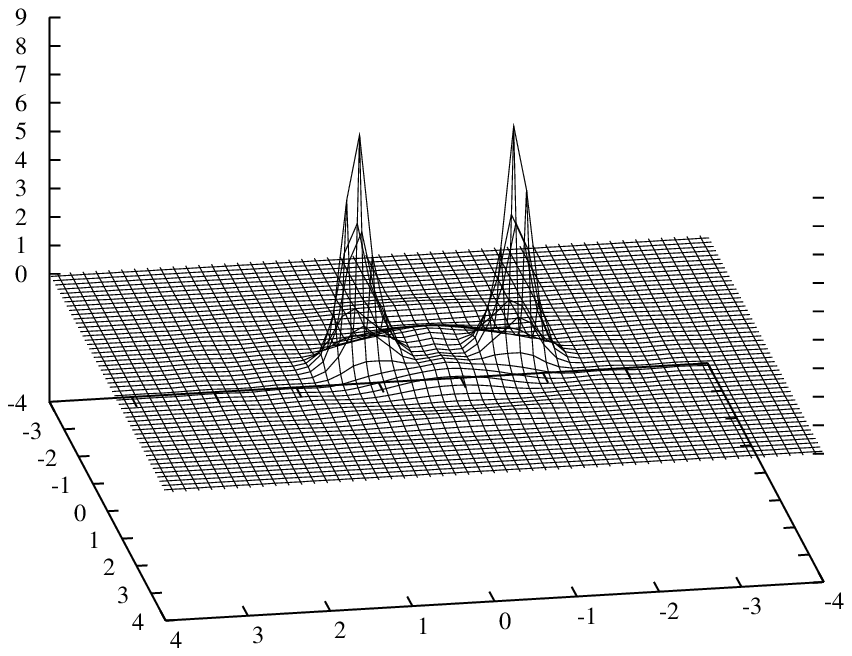,width=10cm}}\\
%%%%%%%%%%
\raisebox{3.4cm}[0.5cm][0cm]{
\hspace{4.0cm}$|\Psi_2|^2$}\\ 
\raisebox{0.9cm}[0cm][0cm]{
$\hspace{4.8cm}\t$}\\
\raisebox{2.9cm}[0cm][0cm]{
\hspace{1.5cm}b)}\\
\raisebox{0.9cm}[0cm][0cm]{
\hspace{3.5cm}\hspace{5.1cm}$\xi$}\\
\raisebox{2.4cm}[2.0cm][1cm]{
\hspace*{3.5cm}\psfig{figure=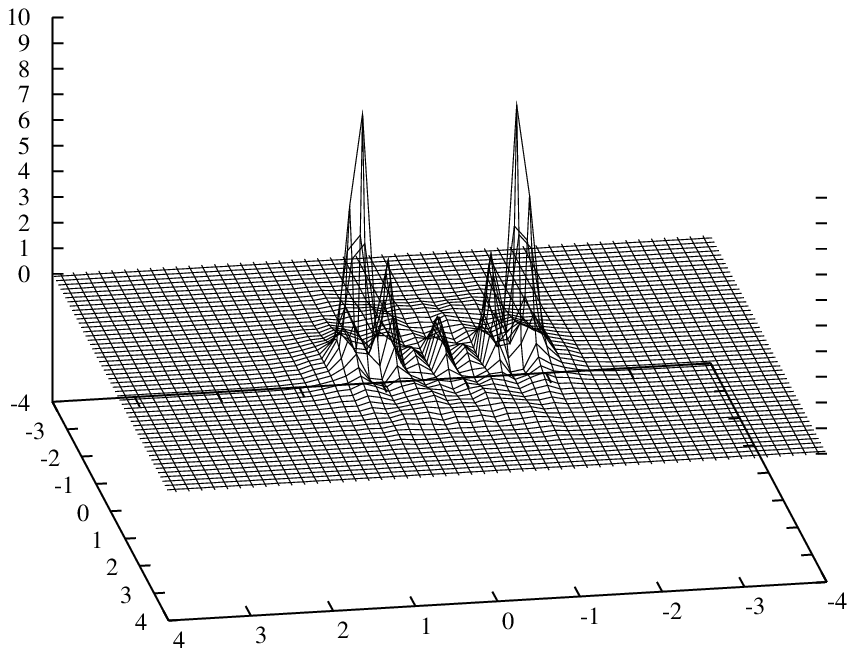,width=10cm}}
%%%%%%%%%%
\raisebox{3.8cm}[0.5cm][0cm]{
\hspace{4.0cm}$|\Psi|^2$ }\\
\raisebox{1.3cm}[0cm][0cm]{
$\hspace{4.8cm}\t$}\\
\raisebox{3.3cm}[0cm][0cm]{
\hspace{1.5cm}c)}\\
\raisebox{1.3cm}[0cm][0cm]{
\hspace{3.5cm}\hspace{5.1cm}$\xi$}\\
\raisebox{2.8cm}[2.0cm][1cm]{
\hspace*{3.5cm}\psfig{figure=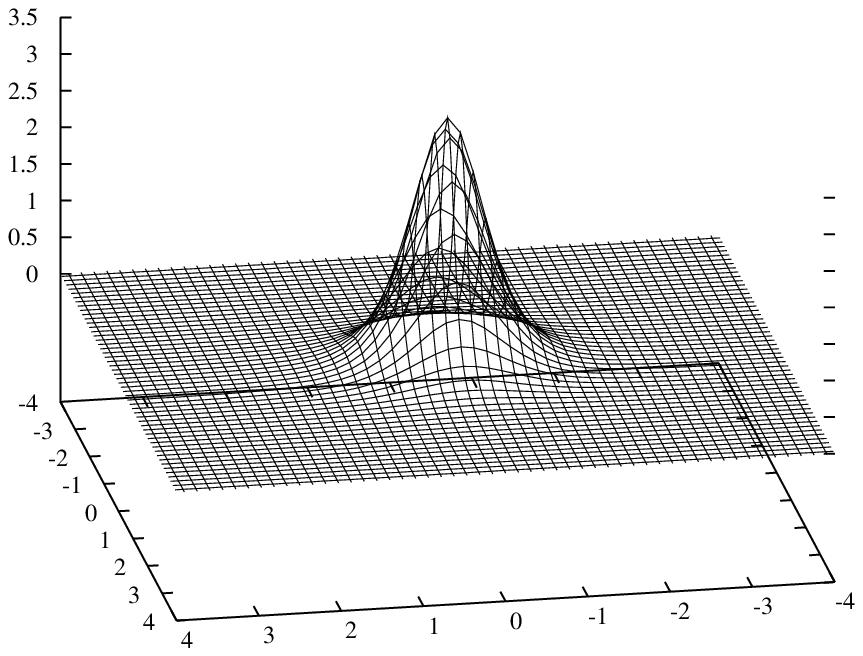,width=10cm}}\\
%%%%%%%%%%
\raisebox{4.0cm}[0.5cm][0cm]{
\hspace{4.0cm}$|\Psi|^2$ }\\
\raisebox{1.5cm}[0cm][0cm]{
$\hspace{4.8cm}\t$}\\
\raisebox{3.5cm}[0cm][0cm]{
\hspace{1.5cm}d)}\\
\raisebox{1.5cm}[0cm][0cm]{
\hspace{3.5cm}\hspace{5.1cm}$\xi$}\\
\raisebox{3.0cm}[2.0cm][1cm]{
\hspace*{3.5cm}\psfig{figure=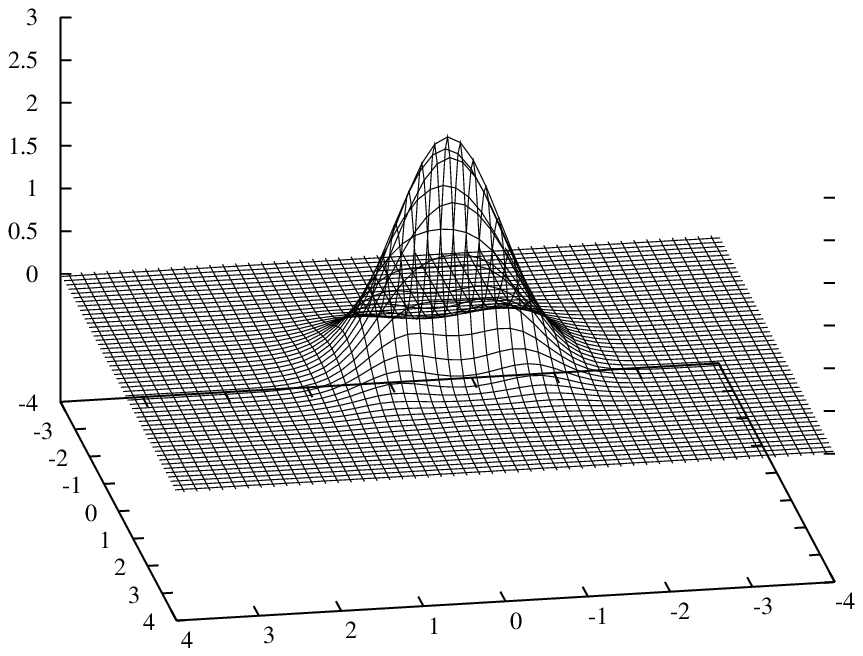,width=10cm}}
\raisebox{3.5cm}[0cm][0cm]{
\hspace{8cm}$\z=1.0$}
\vspace*{-3.6cm}
\caption{The spatio-temporal dependence of the intensity of the anomalous 
pulse (a) co-propagating with the normal pulse (b) and the spatio-temporal 
dependence of the intensities of both pulses, anomalous one (c) and normal 
one (d) when they propagate separately. The following parameters of the
system have been chosen: $\sigma_1=1.0, \k_1=1.88, \sigma_2=-0.1, 
\k_2=2.0$; the propagation distance $\z=1.0$.}
\label{figdistl}
\end{figure}

\begin{figure}
\raisebox{1.cm}[2.5cm][0cm]{
\hspace{4.0cm}$|\Psi_1|^2$}\\ 
\raisebox{-1.5cm}[0cm][0cm]{
$\hspace{4.8cm}\t$}\\
\raisebox{0.5cm}[0cm][0cm]{
\hspace{1.5cm}a)}\\
\raisebox{-1.5cm}[0cm][2cm]{
\hspace{3.5cm}\hspace{5.1cm}$\xi$}\\
\raisebox{2cm}[2.0cm][1cm]{
\hspace*{3.5cm}\psfig{figure=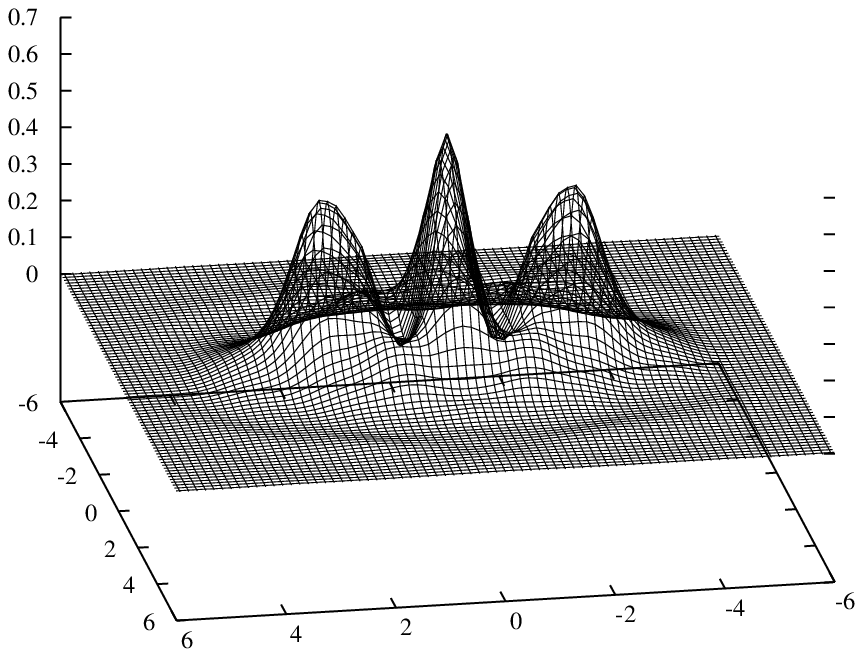,width=10cm}}\\
%%%%%%%%%%
\raisebox{3.4cm}[0.5cm][0cm]{
\hspace{4.0cm}$|\Psi_2|^2$}\\ 
\raisebox{0.9cm}[0cm][0cm]{
$\hspace{4.8cm}\t$}\\
\raisebox{2.9cm}[0cm][0cm]{
\hspace{1.5cm}b)}\\
\raisebox{0.9cm}[0cm][0cm]{
\hspace{3.5cm}\hspace{5.1cm}$\xi$}\\
\raisebox{2.4cm}[2.0cm][1cm]{
\hspace*{3.5cm}\psfig{figure=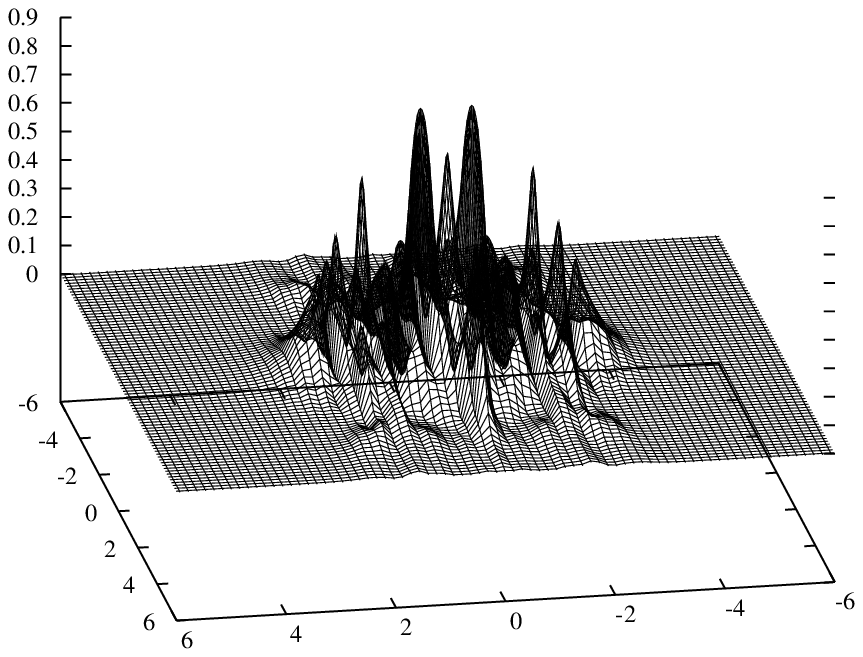,width=10cm}}
%%%%%%%%%%
\raisebox{3.8cm}[0.5cm][0cm]{
\hspace{4.0cm}$|\Psi|^2$ }\\
\raisebox{1.3cm}[0cm][0cm]{
$\hspace{4.8cm}\t$}\\
\raisebox{3.3cm}[0cm][0cm]{
\hspace{1.5cm}c)}\\
\raisebox{1.3cm}[0cm][0cm]{
\hspace{3.5cm}\hspace{5.1cm}$\xi$}\\
\raisebox{2.8cm}[2.0cm][1cm]{
\hspace*{3.5cm}\psfig{figure=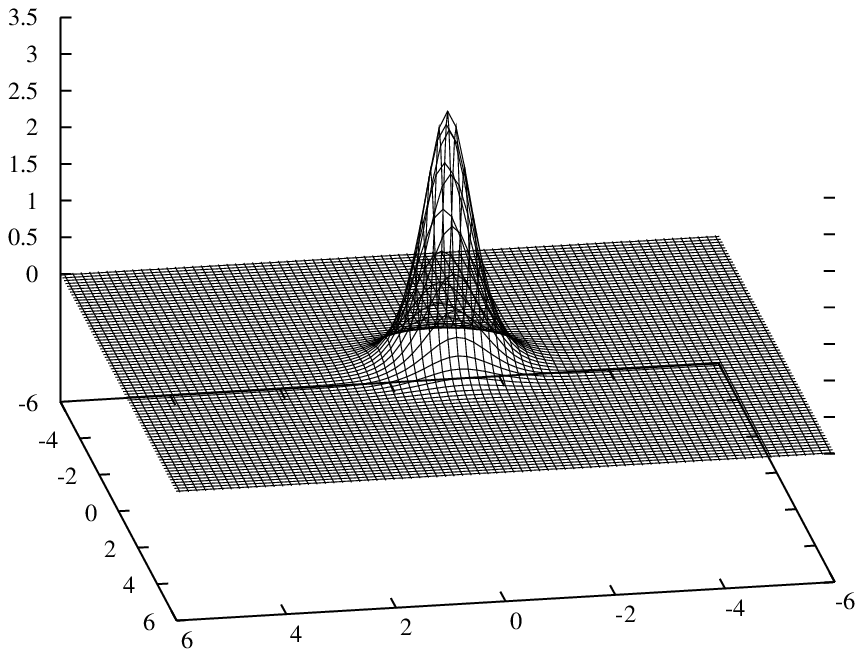,width=10cm}}\\
%%%%%%%%%%
\raisebox{4.0cm}[0.5cm][0cm]{
\hspace{4.0cm}$|\Psi|^2$ }\\
\raisebox{1.5cm}[0cm][0cm]{
$\hspace{4.8cm}\t$}\\
\raisebox{3.5cm}[0cm][0cm]{
\hspace{1.5cm}d)}\\
\raisebox{1.5cm}[0cm][0cm]{
\hspace{3.5cm}\hspace{5.1cm}$\xi$}\\
\raisebox{3.0cm}[2.0cm][1cm]{
\hspace*{3.5cm}\psfig{figure=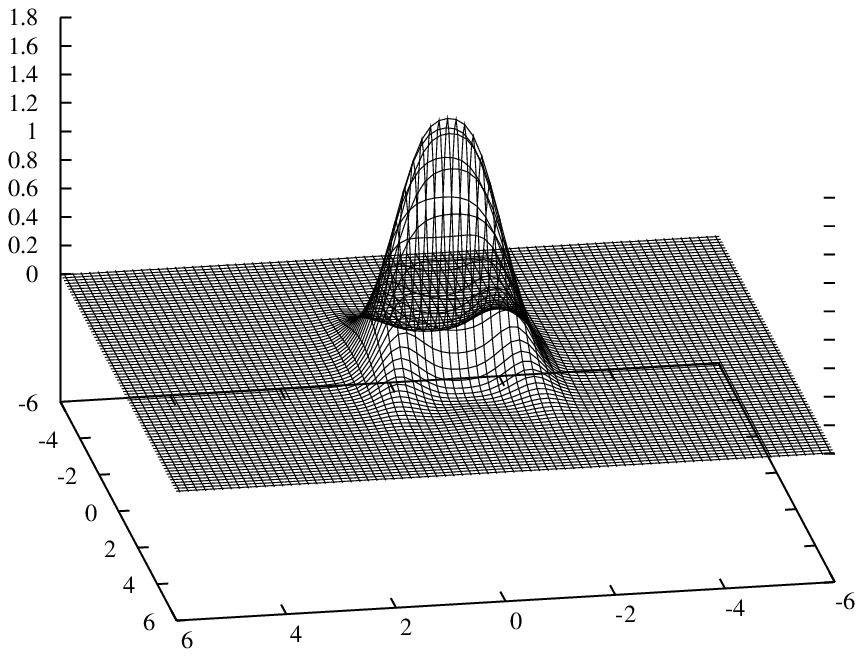,width=10cm}}
\raisebox{3.5cm}[0cm][0cm]{
\hspace{8cm}$\z=1.0$}
\vspace*{-3.6cm}
\caption{The spatio-temporal dependence of the intensity of the anomalous 
pulse (a) co-propagating with the normal pulse (b) and the spatio-temporal 
dependence of the intensities of both pulses, anomalous one (c) 
and normal one (d) when they propagate separately. The parameters
of the system are the same as in the case considered in figure 3:
%\ref{figdistl}: 
i.e., $\sigma_1=1.0, \k_1=1.88, \sigma_2=-0.1, \k_2=2.0$; 
the only difference is that the distance of propagation is larger, 
$\z=2.0$.}
\label{figdistr}
\end{figure}

Since the variational method is not applicable to the study of 
spatio-temporal splitting of two pulses propagating
simultaneously in a nonlinear planar waveguide, the results 
of this section are due to numerical simulations.
Figures \ref{figdistl} and \ref{figdistr} show the spatio-temporal 
dependences of intensities of both pulses, the anomalous one 
(figure \ref{figdistl}(a) and \ref{figdistr}(a)), and the normal one 
(figure \ref{figdistl}(b) and \ref{figdistr}(b)), for different 
longitudinal variables, $\z$. Parameters of the pulses were chosen 
in such a way that when they propagate as single pulses the following 
effects take place on large propagation distances: (i) symmetric, 
spatio-temporal broadening of the anomalous pulse without occurence of 
catastrophic self-focusing (see figures \ref{figdistl}(c) and 
\ref{figdistr}(c)); and (ii) large asymmetrical, spatio-temporal 
broadening of the normal pulse without splitting into sub-pulses 
(see figures \ref{figdistl}(d) and \ref{figdistr}(d)). 
The case (i) occurs when the conditions $\s_1=1$ and 
$\k_1< \k_{V cat} = 1+\s_1$ are satisfied, while the case (ii) does when
the strength of nonlinearity, $\k_2$, is sufficiently small. 
When the pulses propagate simultaneously, i.e. there is a nonlinear 
coupling between them, the situation becomes qualitatively different,
as it can be seen from figures \ref{figdistr}(a) and \ref{figdistr}(b).  
Namely, spatio-temporal splitting of both pulses can develop, so that
for the propagation distance $\z=2$ the anomalous (normal) pulse 
becomes divided into $n=3$ $(n>10)$
sub-pulses. The effect of splitting of the anomalous pulse, which does not
occur when it propagates as a single pulse, can be explained as follows.  
When the nonlinear coupling between pulses through cross-phase modulation 
is present one pulse can induce a redistribution of energy of the other 
pulse. Therefore, if there are some local focusing areas in the 
distribution of energy of one pulse, energy of the other pulse 
tends to concentrate there. Such a tendency has been already pointed 
out by Agrawal, who has observed the occurrence of local focusing 
areas in the distribution of energy of two beams which co-propagate 
in a defocusing nonlinear medium \cite{Agrawal:ifo}.

\section{The limiting case of vanishing dispersion of the normal pulse}
In this section we consider the limiting case when the dispersive term 
of the normal pulse can be neglected. We apply the variational 
method and numerical simulations and compare 
their results. We assume that the initial condition has the 
shape of the Gaussian function given by equation (\ref{gauss}) and 
concentrate basically  on the question as to whether there exists
a stable self-trapped solution of the above-mentioned system of equations. 

First we briefly discuss the case when the pulses propagate in a planar 
waveguide se\-pa\-ra\-te\-ly, i.e. when there is no coupling between them. 
Specifically, we consider (i) the propagation of a pulse with anomalous 
dispersion and (ii) the propagation of a dispersion-less beam. 
The case (i)  can be described by the (2+1)-dimensional NSE, 
which does not have stable, self-trapped solutions. Thus, depending 
on parameters of the system, either spatio-temporal spreading of 
the pulse or catastrophic self-focusing develops. The case (ii) is 
described by the (1+1)-dimensional NSE which depends only 
on one transverse variable $\xi$ and being an integrable system possesses 
the familiar soliton solution given by $sech$ function \cite{Zakharov:eto}. 
Taking the Gaussian function (equation (\ref{gauss})) which depends on two 
transverse variables $\t$ and $\xi$  from the variational method we 
obtain that the temporal width of the pulse is constant while the 
spatial width oscillates. These oscillations are due to the fact that 
the shape of the Gaussian trial function differs from the exact 
soliton solution given by the $sech$ function \cite{Anderson:vat}. 
However, numerical simulations lead to a slightly different behavior. Namely, 
the temporal width of the pulse appears to oscillate in sinchronization 
with the spatial width. Amplitudes of both oscillations decrease with 
the longitudinal variable $\z$ and vanish at finite $\z$ when the 
spatial soliton is formed \cite{Burak:gbp}.
 
Now, let us take into account the nonlinear coupling between pulses
and discuss briefly the aspect of catastrophic self-focusing. 
Based on the variational method we have observed that when 
parameters of the anomalous pulse are chosen in such a way that 
catastrophic self-focusing does not occur when it propagates 
as a single pulse, i.e. when the condition $\k_1 < \k_{V cat} = 1 + \s_1$
is satisfied, then also in the case of the anomalous pulse coupled to the 
normal pulse no catastrophic self-focusing of both of them occurs, 
even for very large strength of nonlinearity of the normal pulse, i.g.,
$\k_2=20$. However, when the condition $\k_1 < \k_{V cat} = 1 + \s_1$ 
is not fulfilled a similarity to the case discussed in section 3 can 
be found, namely three different behaviours of the pulses can be observed:
(i) no catastrophic self-focusing of both pulses; (ii) catastrophic 
self-focusing of the anomalous pulse; (iii) catastrophic self-focusing 
of both pulses, the spatial width of the normal pulse vanishes to 
zero while his temporal width remains larger than zero. 

The above results obtained in the variational method have been verified 
in the numerical simulations except for the case (iii): that is a 
development of catastrophic self-focusing of the normal pulse whose
spatial width venishes to zero on a certain distance of propagation, $\z$,  
and whose temporal width is left larger than zero has not been observed,
but also no definite statement that this effect is prohibited can be made.
Rather, we are interested in possibilities of formation of self-trapped 
solutions and, therefore, we will restrict our analysis to parameters 
of the pulses which assure that no catastrophic self-focusing occurs,
i.e., that condition $\k_1 < \k_{N cat} \approx 0.885 + \s_1$ is satisfied.

\begin{figure}
\raisebox{-2.5cm}[1.0cm][1cm]{
a)}\\
\hspace*{1cm}\raisebox{-2cm}[4cm][2cm]{
\psfig{figure=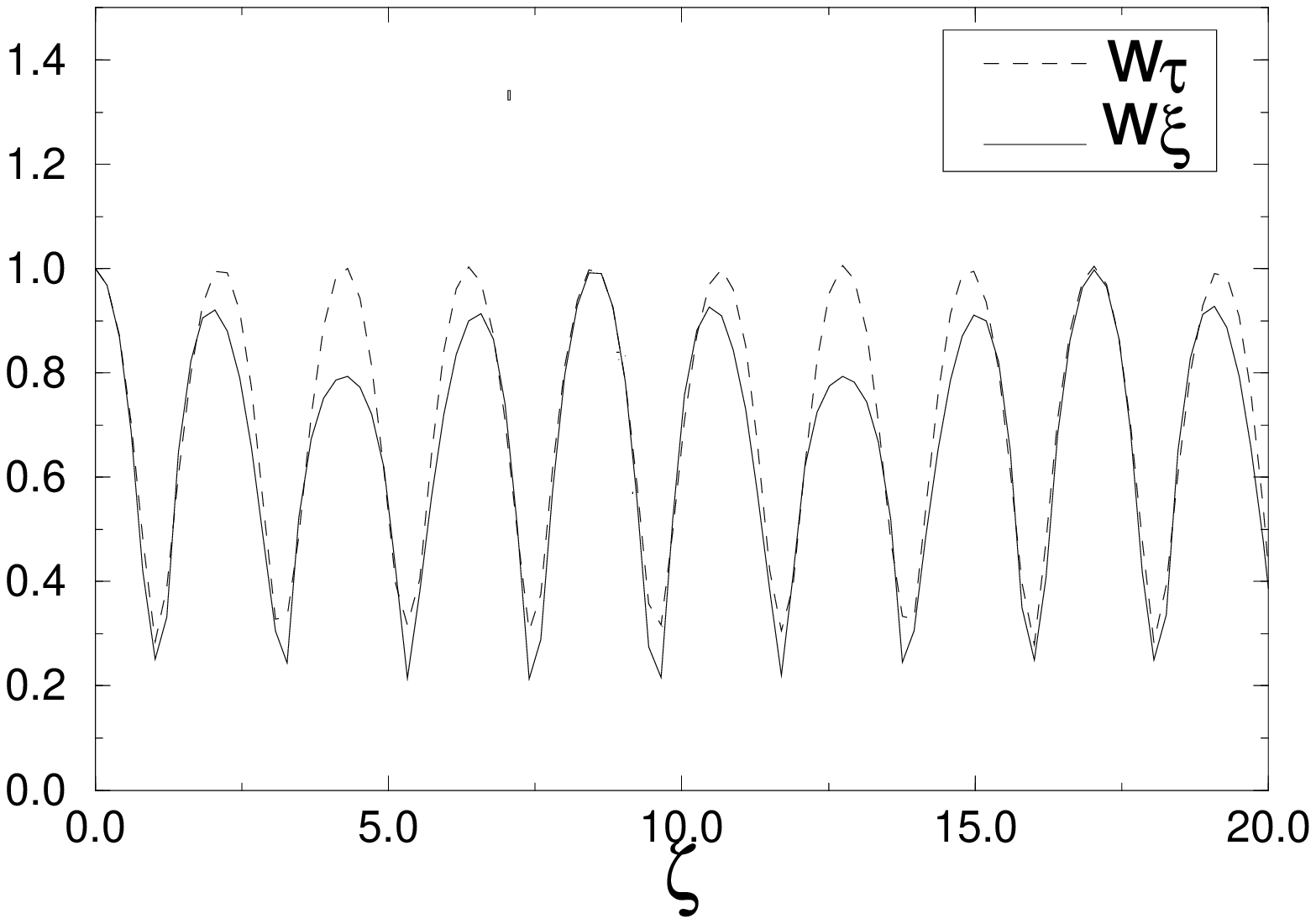,width=13cm}}\\
\raisebox{-2.5cm}[1cm][1cm]{
b)}\\
\hspace*{1cm}\raisebox{-2cm}[4cm][2cm]{
\psfig{figure=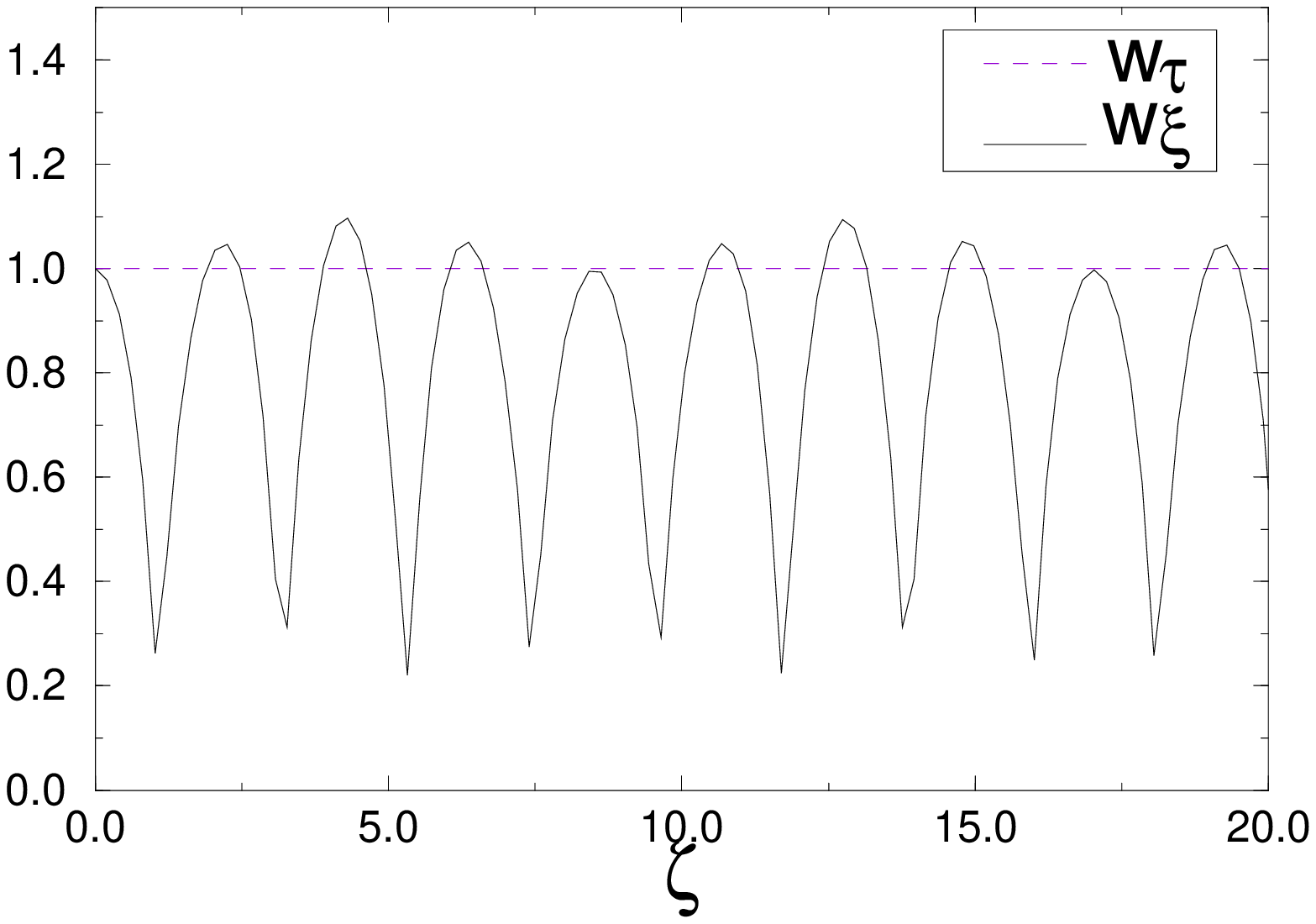,width=13cm}}
\caption{The results obtained using the variational method displaying the
dependence of the temporal, $w_{\t}$, and spatial, $w_{\xi}$, widths of the 
anomalous pulse (a) co-propagating with the normal pulse (b), 
for the following parammeters of the system:
$\k_1 = 1.0, \s_1=1.0, \k_2=2.0, \s_2=0.0$.}
\label{figv}
\end{figure}

From the variational method it follows that the evolution of the normal 
pulse coupled to the anomalous one is essentially similar to that of 
the single normal pulse. Namely, the temporal width of the pulse 
does not depend on the longitudinal variable, $\z$, as is seen from 
equation (8c) with the neglected dispersion of the normal pulse $\s_2=0$,
while the spatial width of the pulse undergoes periodic oscillations
(see figure \ref{figv}(b)). The propagation of the anomalous pulse 
coupled to the normal one is, however, qualitatively different to 
the behaviour of a single anomalous pulse. Namely, both temporal 
and spatial widths of the pulse undergo periodic oscillations 
(see figure \ref{figv}(a)). Therefore, neither spatio-temporal spreading 
nor catastrophic self-focusing of the anomalous pulse can develop and 
a self-trapped solution arises. Note that a similar self-trapped solution 
was found in the case of (2+1)-dimensional NSE with the saturation 
of nonlinearity \cite{Karlsson:obi}. 
 
\begin{figure}
\raisebox{-2.5cm}[1.0cm][1cm]{
a)}\\
\hspace*{1cm}\raisebox{-2cm}[4cm][2cm]{
\psfig{figure=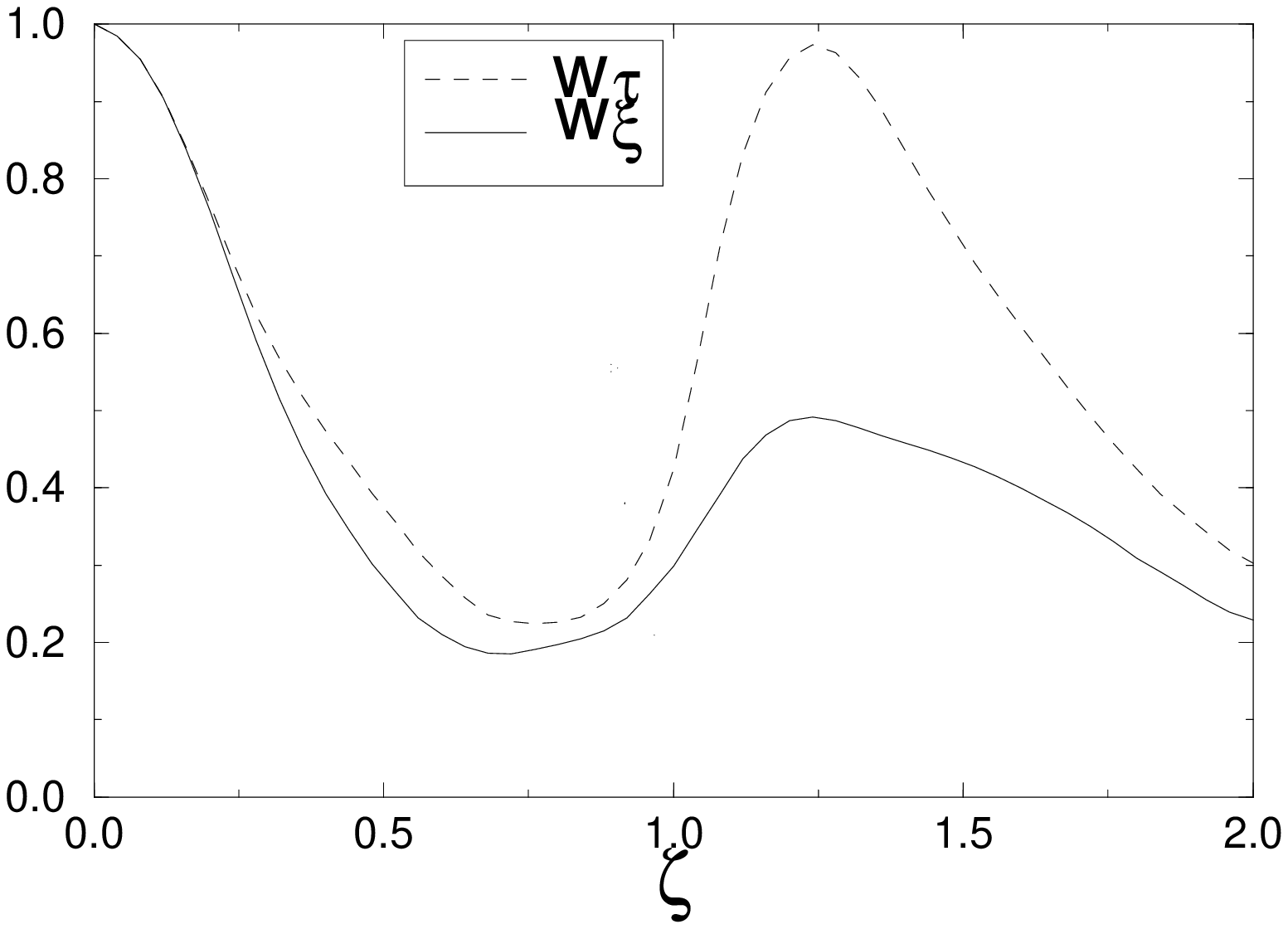,width=13cm}}\\
\raisebox{-2.5cm}[1cm][1cm]{
b)}\\
\hspace*{1cm}\raisebox{-2cm}[4cm][2cm]{
\psfig{figure=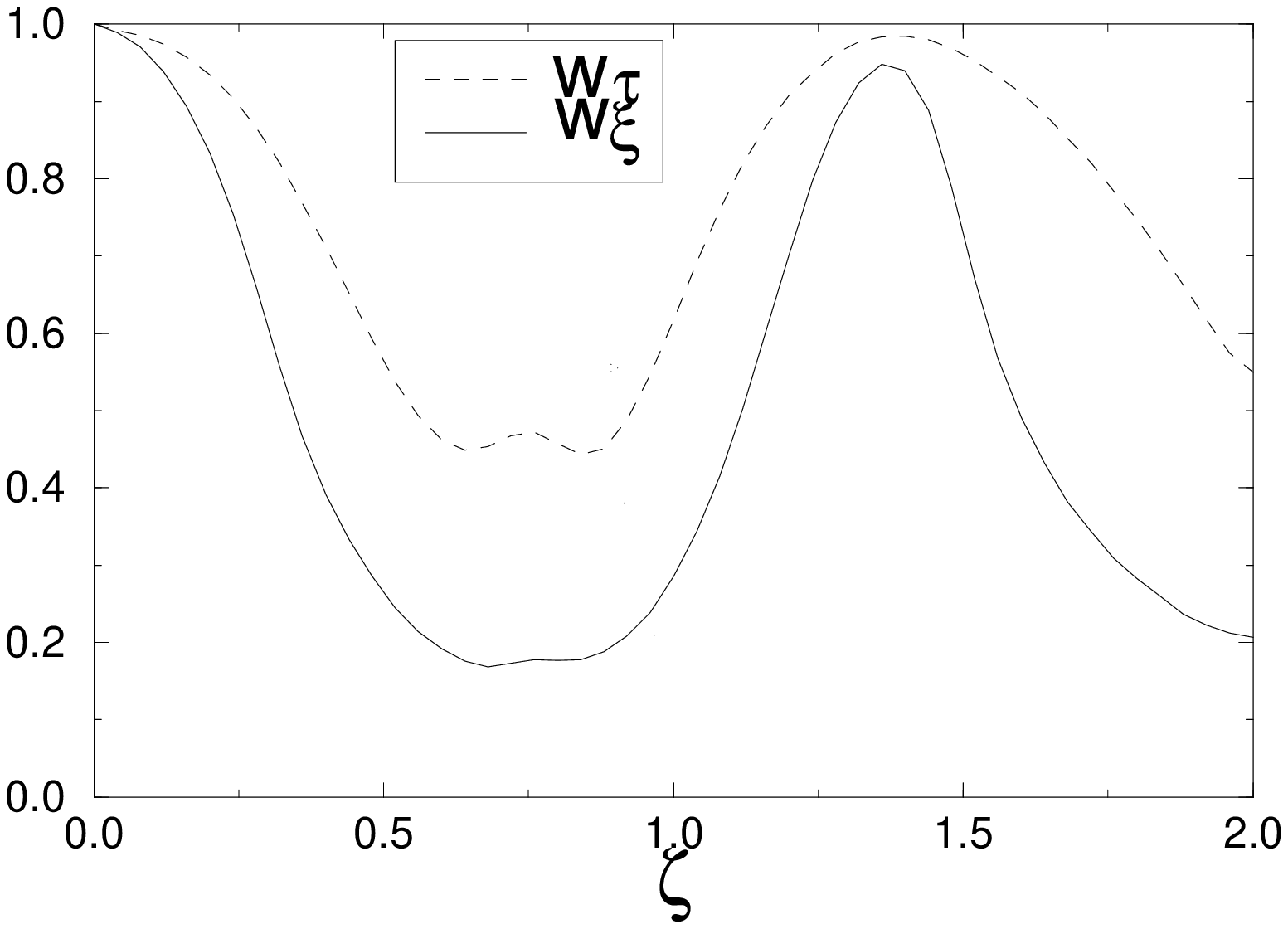,width=13cm}}
\caption{The results obtained using the numerical simulations displaying the
dependence of the temporal, $w_{\t}$, and spatial, $w_{\xi}$, widths of the 
anomalous pulse (a) co-propagating with the normal pulse (b)
for the following parammeters of the system:
$\k_1 = 1.0, \s_1=1.0, \k_2=2.0, \s_2=0.0$.}
\label{fign}
\end{figure}

We also performed  numerical simulations for the case of simultaneously
propagating pulses. The results are displayed in figure \ref{fign}  
from which it is evident that the temporal and spatial widths of both 
pulses oscillate in sinchronization, with the amplitude of the temporal 
oscillations smaller than the amplitude of the spatial ones. 
Unfortunately, the numerical calculations are rather labourious and 
we have not yet been able to calculate evolution for longer longitudinal 
variables, $\z > 2$, so that we do not know whether the amplitude of 
oscillations decreases with $\z$ and whether or not no spreading and 
catastrophic self-focusing of the anomalous 
pulse develop. Nevertheless, the currently available numerical 
results suggest that a self-trapped solution can exist in the 
configuration under discussion. Further calculations should clarify 
this question. 

Note that a configuration of two simultaneously propagating pulses 
could also be used in optical compression techniques since, 
as is seen  from figure \ref{fign}(a), for some particular values of 
the longitudinal distance $\z$ the temporal width of the anomalous pulse 
decreases by about five times the initial width. 
 
\section{Conclusions}
In this paper properties of two pulses propagating simultaneously
in different dispersion regimes, i.e. anomalous and normal, in a Kerr-type 
planar waveguide are considered. The propagation is described 
by two coupled NSEs. The interaction between pulses is assumed to be 
limited to cross-phase modulation. Four wave mixing is neglected, 
i.e. no energy transfer between pulses is taken into account. 
The accuracy of another assumption used in the analysis, 
the omitting of the difference of group velocities of the pulses, 
is discussed in appendix B. Our analysis is based 
on the variational method and numerical simulations.

First we studied  the influence of the parameters of the pulse 
propagating in a normal dispersion regime on the threshold of catastrophic 
self-focusing of the pulse with an anomalous dispersion.
We observed that catastrophic self-focusing of the pulse propagating 
in an anomalous dispersion regime can be arrested by the pulse propagating 
in a normal dispersion regime when the strength of nonlinearity is 
sufficiently large, $\k_2 > \k_{X low}$ and the dispersion-to-diffraction 
ratio satisfies the relation: 
$|\s_{X low}(\k_2)| < |\s_2| < |\s_{X upp}(\k_2)|$. In this notation
$X \equiv V (X \equiv N)$ concerns the results obtained in the 
variational method (numerical simulations). 
We also investigated whether the nonlinear coupling between pulses can 
cause catastrophic self-focusing of the pulse propagating in a normal 
dispersion regime. The variational method indicates that when 
catastrophic self-focusing of the anomalous pulse occurs, the normal 
pulse can display, depending on the parameters of the system, 
two different characteristics: (i) both widths of the pulse initially 
decrease reaching a minimum on a certain distance of propagation 
and then they start to increase, (ii) the spatial width of the pulse 
vanishes to zero on a finite distance of propagation whereas the
temporal width initially decreases reaching a minimum on a certain 
distance of propagation and then it increases. The occurence
of catastrophic self-focusing of the normal pulse has not been observed
in the numerical simulations. Therefore, we can conclude,
based on the variational method and the numerical simulations, 
that catastrophic self-focusing of the anomalous pulse does not 
necessarily lead to catastrophic self-focusing of the normal pulse. 
 
We found also, using the numerical simulations, that the presence of the
pulse propagating in a normal dispersion regime can lead to spatio-temporal
splitting of the pulse propagating in an anomalous dispersion regime. 
Recall that splitting of an anomalous pulse into several pulses 
does not occur when it propagates as a single pulse. 

Finally, we considered the limiting case of vanishing dispersion of the 
pulse propagating in a normal dispersion regime with
parameters of the pulses chosen in such a way that catastrophic 
self-focusing does not occur, i.e. that the conditions 
$\k_1 < \k_{V cat} = 1 + \s_1$ (in the variational method) and
$\k_1 < \k_{N cat} \approx 0.885 + \s_1$ (in the numerical simulations) 
are satisfied. The main motivation 
was to see whether such a configuration can lead to a stable self-trapped
propagation of a pulse with anomalous dispersion. The positive answer was
obtained within the variational method 
which confirms that neither spatio-temporal spreading nor catastrophic 
self-focusing of the a\-no\-ma\-lo\-us pulse can develop 
thus giving rise to a self-trapped solution.
Note that this kind of stabilization is similar to that which 
has been found earlier in media with saturation-type nonlinearity 
\cite{Karlsson:obi}. Although the existing data supports the existence 
of a self-trapped solution, conclusive results require labourious 
simulations at high values of the longitudinal variable $\z$ and are 
not yet available (work in progress). 

Note, in conclusion, that the existence of a stable self-trapped solution 
could be useful, for example, in optical switching devices. 
The configuration of two simultaneously propagating pulses in a planar 
waveguide could also be of use in optical compression techniques. 

\section{Acknowledgements}
The work was supported by the Polish Committee of Scientific Research 
(KBN, grant no 8T11F 007 14) and the Deutsche Akademische Austauschdienst 
(DAAD), to both of which I express my gratitude. I take the opportunity 
to express my thanks to Professor F. Lederer for his kind hospitality at
the Institute of Solid State Physics and Theoretical Optics, 
Fredrich-Schiller-Universit\"at Jena, Jena, Germany. The numerical 
calculations were partially done thanks to a fellowship at the Abdus 
Salam International Centre for Theoretical Physics, Trieste, Italy. 
I gratefully acknowledge the Director of the Centre Professor M. Virasoro, 
and Professor G. Denardo for their kind hospitality and helpful support. 
I also would like to thank the referees for constructive 
suggestions.  

\subsection*{Appendix A}
The notation in equations (\ref{nse}) and (\ref{2nse}a,b) 
is as follows \cite{Ryan:pca}:  
$\zeta = z / {z_{DF1}}$ is the longitudinal coordinate
normalized to the Fresnel diffraction length of the anomalous 
pulse, $\xi = x / {w_1}$ is the spatial transverse coordinate
normalized to the initial spatial width of the anomalous pulse, 
$\tau = ({t-\beta _1^{(1)} z})/{t_1}$ is the local time normalized to 
the initial temporal width of the anomalous pulse. The parameters
$\sigma_j = {z_{DF1}}/{z_{DSj}}$, $\mu={z_{DF1}}/{z_{DF2}}$, 
$r={\lambda_1}/{\lambda_2}= {\omega_2}/{\omega_1}$
denote, respectively, the dispersion-to-diffraction ratio, 
the ratio of the Fresnel diffraction length of the anomalous pulse 
to the Fresnel diffraction length of the normal pulse and, finally,
the ratio of the carrier frequency of the anomalous pulse
to the carrier frequency of the normal pulse. 
$\Psi_j:= \sqrt{\k_j} {U_j(\z, \t, \xi)}/{U_{j 0}}$
denotes the normalized amplitude of the $j$-s pulse, where 
$U_j(\z, \t, \xi)$ is the amplitude of the slowly varying envelope of 
the electric field, $U_{j 0} := U_j(0,0,0)$ is the dimensionless 
initial peak amplitude. The parameter 
$ \k_j := \left( {z_{DF j}} / {z_{NL j}} \right)^2$ defined as the
strength of nonlinearity of j-s pulse is proportional 
to the nonlinear part, $n_2$, of the refractive index of a medium, 
$n:=n_0 + n_2 |U_j|^2$, to the initial peak intensity, 
$|U_{j 0}|^2$, and to the square of the spatial 
width of the pulse, $w^2_{\xi j}$. Note also that in the case of the 
(1+1)-dimensional NSE, i.e. when $\s_j=0$, the quantity 
$\sqrt{\k_j}$ can be interpreted as the order of a spatial soliton, 
so that a first-order soliton arises when $\k_j=1$ 
\cite{Agrawal:nfo}. The dispersive terms are defined as follows:
$\beta_j^{(0)}:=\beta^{(0)}(\omega_j)= {\omega_j}/ c$ is the wavenumber, 
$\beta_j^{(1)}:={\pr \beta}/{\pr \omega}|_{\omega=\omega_j}=1 /v_{gj}$
is the reverse group velocity, and
$\beta_j^{(2)}:= {\pr^2 \beta}/{\pr \omega^2}|_{\omega=\omega_j}$
is the group velocity dispersion. The parameters
$z_{DFj} := \beta_j^{(0)} n_0(\omega_j) w_{i}^2$,
$z_{DSj}:= {t_j^2}/{\beta_j^{(2)}}$, 
$z_{NLj}:= w_j \sqrt{{n_0}/{(2 n_2 |U_{ 0j}|^2)}}$,
$w_j$, $t_j$ denote, respectively, 
the Fresnel diffraction length, the dispersive length, the nonlinear length,
the initial spatial width and the initial temporal width of the $j$-s pulse.
In the above notation $j=1,2$, where the subscript $j=1$ ($j=2$) 
refers to the anomalous (normal) pulse.

\subsection*{Appendix B}
Since we have assumed that pulses have different wavelengths and 
different group velocity dis\-per\-sions, it is physically evident that 
they should also have different group velocities. Therefore, the 
assumption that the difference of the group velocities of the pulses 
vanishes is a simplification accepted in this paper and should be 
viewed as a first step of the analysis. When this difference does not 
vanish the pulses propagate with different velocities and the overlap 
between them decreases with the longitudinal variable. Therefore, 
the nonlinear coupling between them also decreases. In the limiting case 
of the difference of the group velocities of the pulses approaching 
infinity the coupling between pulses becomes zero and the problem of 
simultaneous propagation of two pulses reduces to the case when they 
propagate separately. 

However, we believe that the inclusion of a small difference of group 
velocities of the pulses, which should be studied numerically, 
will not cause qualitative changes in the 
results of this paper, such as the possibility of an arresting of 
catastrophic self-focusing of the pulse propagating in an anomalous 
dispersion regime by the influence of the pulse propagating in a normal 
dispersion regime. The only difference we expect is a change of the 
values of the parameters, $\s_{N low}, \s_{N upp},$ $\k_{N upp}$, 
which describe the threshold of catastrophic self-focusing for 
fixed values of $\s_1$ and $\k_1$. These quantitative changes 
would be proportional to the value of the difference of the group 
velocities of the pulses. 

%\section*{References}

\small
\bibliography{papers}

\end{document}